\documentclass[acmsmall,screen,nonacm]{acmart}

\AtBeginDocument{%
  }

\setcopyright{acmcopyright}
\copyrightyear{2018}
\acmYear{2018}
\acmDOI{XXXXXXX.XXXXXXX}

\acmConference[Conference acronym 'XX]{Make sure to enter the correct
  conference title from your rights confirmation emai}{June 03--05,
  2018}{Woodstock, NY}
\acmPrice{15.00}
\acmISBN{978-1-4503-XXXX-X/18/06}

\usepackage{titlesec}
\usepackage{lipsum}
\usepackage{multicol}

\usepackage{siunitx}
\usepackage{datapie}
\usetikzlibrary{patterns}

\usepackage{algorithmicx} 
\usepackage{algpseudocode} 

\newcommand\cmt[1]{{\scriptsize\ttfamily\textcolor{blue}{// #1\\}}}

\usepackage{microtype}
\usepackage{booktabs}
\usepackage{subcaption}
\usepackage{soul}

\usepackage[boxed,vlined,ruled,algo2e]{algorithm2e}
\SetAlCapSkip{0.5em}

\usepackage{amsmath}
\usepackage{stmaryrd}
\usepackage{enumitem}
\usepackage{wrapfig}
\usepackage{tikz}
\usetikzlibrary{shapes}
\usetikzlibrary{shapes.arrows}
\usetikzlibrary{fit}
\usepgflibrary{shapes.symbols}

\newcommand{\alg}[1]{{\bf \texttt{#1}}\xspace}

\newcommand{\default}{\alg{Default}}  
\newcommand{\gfm}{\alg{GFM}}  
\newcommand{\lfm}{\alg{LFM}}  
\newcommand{\gfo}{\alg{GFO}}  
\newcommand{\lfo}{\alg{LFO}}  

\newcommand{\CodeGen}{\textit{CodeGen}\xspace}
\newcommand{\Opt}{\textit{Opt}\xspace}
\newcommand{\GlobalPrefixTree}{\textit{GlobalPrefixTree}\xspace}
\newcommand{\GlobalMergeInfo}{\textit{GlobalMergeInfo}\xspace}

\newcommand{\fb}{\texttt{SocialApp1}\xspace}
\newcommand{\bizapp}{\texttt{SocialApp2}\xspace}
\newcommand{\fbatwork}{\texttt{SocialApp3}\xspace}
\newcommand{\msg}{\texttt{ChatApp1}\xspace}
\newcommand{\talk}{\texttt{ChatApp2}\xspace}

\usepackage{caption}
\usepackage{pgf-pie}

\titleformat{\paragraph}[runin]
{\bfseries} 
{} 
{0pt} 
{} 

\begin{document}

\title{Optimistic Global Function Merger}

\author{Kyungwoo Lee}
\orcid{0000-0002-9127-7261}             
\affiliation{
	\institution{Meta}            
	\city{Menlo Park, CA}
	\country{USA}                    
}
\email{kyulee@meta.com}          

\author{Manman Ren}
\orcid{0009-0006-7023-9449}             
\affiliation{
	\institution{Meta}            
	\city{Menlo Park, CA}
	\country{USA}                    
}
\email{mren@meta.com}          

\author{Ellis Hoag}
\orcid{0000-0003-3853-1889}             
\affiliation{
  \institution{Meta}            
  \city{Menlo Park, CA}
  \country{USA}                    
}
\email{ellishoag@meta.com}          


\begin{abstract}
Function merging is a pivotal technique for reducing code size by combining identical or similar functions into a single function.
While prior research has extensively explored this technique, it has not been assessed in conjunction with function outlining and linker's identical code folding, despite substantial common ground.
The traditional approaches necessitate the complete intermediate representation to compare functions.
Consequently, none of these approaches offer a scalable solution compatible with separate compilations while achieving global function merging, which is critical for large app development.
In this paper, we introduce our global function merger, leveraging global merge information from previous code generation runs to optimistically create merging instances within each module context independently.
Notably, our approach remains sound even when intermediate representations change, making it well-suited for distributed build environments.
We present a comprehensive code generation framework that can run both the state-of-the-art global function outliner and our global function merger.
These components complement each other, resulting in a positive impact on code size reduction.
Our evaluation demonstrates that when integrating the global function merger with a state-of-the-art global function outliner that is fully optimized with ThinLTO, a further reduction of up to 3.5\% in code size can be attained.
This is in addition to the initial average reduction of 17.3\% achieved through global function outlining for real-world iOS apps, all with minimal extra build time.
\end{abstract}






\maketitle

\section{Introduction}

Mobile apps have become an integral part of our daily lives, serving a vast user base.
As new features and user demands continually emerge, it's essential to reduce app size to accommodate these changes.
Smaller app sizes also lead to faster downloads and quicker app launches, significantly enhancing the user experience \cite{R16}.
A recent study \cite{uber} underscores the profound correlation between app size and user engagement, highlighting the important role that size optimization plays in retaining and engaging a wider audience.
However, the challenges extend beyond user satisfaction.
Mobile app distribution platforms, such as the Apple App Store, impose stringent size limitations for downloads using cellular data connections.
If an app surpasses a particular threshold in size, users are strongly discouraged from receiving crucial updates, including vital security enhancements, unless they have access to a Wi-Fi network.
This limitation underscores the critical importance of maintaining a lean app size to ensure the timely delivery of essential improvements, safeguarding both user data and mobile device security.


\paragraph*{Code size optimization.}
Compiler optimizations play a crucial role in compacting mobile apps.
Beyond their performance enhancements, numerous compiler optimizations are equally valuable for code size reduction \cite{sizeopt}.
They encompass the elimination of redundant and unreachable code segments \cite{dce}, common sub-expression elimination \cite{cse}, partial redundancy elimination \cite{pre}, and constant propagation \cite{constprop}.
However, certain performance-focused techniques, such as loop optimizations \cite{loopopt} and inlining \cite{size-inliner}, require careful design, as they can sometimes hinder size reduction efforts.
Given the critical size requirements and the predominantly I/O-bound nature of mobile apps, most mobile app compilers prioritize size optimization, often employing the minimum size optimization flag (-Oz).
Among various code size optimization techniques, two critical ones are function outlining and function merging.

\paragraph*{State of the art.}
Function outlining finds sequences of identical machine code and replaces them with function calls.
LLVM provides a machine function outliner \cite{machineoutliner}, but its scope is limited to a single module.
A recent study \cite{uber} involving commercial iOS apps utilizes \textit{(Full)} link-time optimization (LTO) to combine entire bitcode modules and repeatedly applies the function outlining process to enhance its efficiency.
However, this came at the cost of significantly increased build times.
Global function outlining \cite{pgo-mobile} offers a more scalable approach by overcoming the constraints of local function outlining, especially in the context of ThinLTO \cite{thinlto}.
In another recent development, a custom linker approach \cite{LFZLS22} eliminates the need for the LTO dependency by directly outlining machine instructions at link time.
These function outlining techniques have proven highly effective in reducing application size and have gained widespread adoption in production environments.

While function merging has a similar goal, it targets entire similar functions rather than identical straight-line code sequences.
Depending on language features, such as templates or lambdas in traditional C++ or generics, protocols, and value witness in Swift, many functions exhibit similarities but aren't entirely identical.
Although the function outliner can effectively outline subsets of identical code sequences in these similar functions, it's often more desirable to merge these entire functions, rather than having numerous small outlined sequences.
However, unlike function outlining, function merging is seldom enabled in production environments for several reasons.

\paragraph*{Limitation.}
A function merger \cite{mergefunc} is already available in LLVM, but it can only match identical functions.
The efficiency of the function merger is diminished by the linker's identical code folding (ICF) \cite{linkericf} which performs a similar task.
Swift's legacy merge function \cite{swiftmerge} can merge similar functions that differ by constant operands, but it is not integrated into the primary LLVM pass for other programming languages.
Significant research efforts \cite{ssafuncmerge,hyfm} have employed sequence alignment algorithms to merge arbitrary pairs of functions.
However, these approaches have not been thoroughly evaluated with the minimum size optimization, that enables the function outlining previously discussed.
Merging significantly different functions into a single function can increase the size overhead, whereas outlining identical segments among them is often more profitable.
Consequently, it is questionable whether these approaches effectively reduce code size for real-world mobile apps.


More critically, all previous function merging techniques require the presence of the whole intermediate representations (IR), as they rely on function comparators that operate on IR.
None of these approaches are viable for separate compilations like ThinLTO or NoLTO when aiming to efficiently achieve function merging for a large-scale production binary without significantly extending build times.

\paragraph*{Overview of our approach.}
To address these shortcomings, we introduce a novel function merging framework that complements state-of-the-art global function outlining \cite{pgo-mobile} (\gfo) while supporting efficient separate compilations.
Our newly developed global function merger (\gfm), leverages global merge information obtained from a prior code generation run and optimistically generates merging instances within each module context.
This global merge information is all about grouping functions that are similar but differ only in constants.
It retains the minimum required parameters to distinguish these functions.
Using this information, each similar function instance is independently transformed into a pair of thunk and parameterized function.
When these parameterized functions across modules turn out to be identical, they are folded by the linker's ICF, achieving size reduction.
One significant advantage of this approach is its soundness in the face of changes in the IR, making it applicable to distributed build environments.
To validate the effectiveness of \gfm, we conducted evaluations using several real-world iOS apps.
In addition to the initial 17.3\% average reduction achieved through \gfo, \gfm further contributes to code size savings of up to 3.5\%, resulting in an overall improvement of 20\%.


In summary, our contributions comprise:
\begin{itemize}
  \item As far as our knowledge extends, we are the first to introduce the optimistic global function merging technique in separate compilation environments.
  \item We present a comprehensive code generation framework that seamlessly runs both \gfo and \gfm. These components complement each other, resulting in a positive impact on reducing the size of mobile apps.
  \item Our research demonstrates the effectiveness of our techniques in reducing code size and provides valuable insights into the statistical characteristics of function merging for future reference.
\end{itemize}

The rest of the paper is organized as follows.
Section \ref{sec:background} provides a background of function outlining and function merging with examples.
Section \ref{sec:algorithm} presents our global function merging algorithm.
Section \ref{sec:implementation} delves into the implementation details and practical considerations.
Section \ref{sec:evaluation} showcases our experimental results, followed by a discussion of related work in Section \ref{sec:relatedwork}.
Finally, we wrap up the paper with our conclusions in Section \ref{sec:conclusion}.

\section{Background}\label{sec:background}
In Section \ref{sec:overview_comp}, we compare the general characteristics of the function outliner and function merger, summarized in Table \ref{table:comp}.
Section \ref{sec:globaloutliner} presents an overview of \gfo \cite{pgo-mobile}, which inspired \gfm.
In Section \ref{sec:example}, we provide a motivating example demonstrating the potential for merging functions across different modules, the focus of our global function merger.

\subsection{Function Outliner vs. Function Merger}\label{sec:overview_comp}

\begin{table}[!tb]
  \centering
  \caption{Comparison of Function Outliner and Function Merger in LLVM.}
  \label{table:comp}
  \begin{tabular}{@{} lcc @{}} 
      \toprule
  Category &  Function Outliner & Function Merger \\
  \midrule
  Pass & CodeGen (MIR) & Opt (IR) \\
  Scope &  Block  & Function \\
  Match &  Identical & Similar \\
  Call Overhead & Low & High \\
  Size Saving & High & Medium \\
  Scalable Build & Yes & No \\
  \bottomrule
  \end{tabular}
\end{table}


\paragraph*{Function outliner.}

Function outlining is a key technique in reducing code size by replacing repeated sequences of instructions with function calls.
Within LLVM, two outliners are available: one operating at the LLVM IR level and another at the code generation pass, as described in \cite{machineoutliner}.
Recent studies with commercial mobile apps \cite{uber,pgo-mobile} has shown that outlining at the machine IR (MIR) level in LLVM saves code size more effectively in mobile apps because straight-line short code sequences are heavily repetitive, which are often exposed only after the high-level IR is lowered.
Henceforth, the term \textit{function outliner} in this context refers to the latter case.

The function outliner finds identical straight-line code sequences within a basic block across different functions.
This process occurs after register allocation, when registers are fully assigned.
Consequently, the call to the outlined function typically takes the form of a direct, parameterless call without creating a new frame.
This results in low call overhead.
Due to the frequent reoccurrence of these short code sequences, the reduction in code size achieved through function outlining is substantial.

The default function outliner \cite{machineoutliner} operates within a single module.
Expanding it to cover an entire application typically involves approaches like \textit{(Full)}LTO \cite{uber}, but this can be impractical for production environments due to high compilation costs.
In contrast, \cite{pgo-mobile} introduced \gfo, which utilizes a global summary to outline functions across modules.
This approach supports separate compilations in parallel, operating in the context of ThinLTO.
We will explore this concept in more detail in Section \ref{sec:globaloutliner}.

More recently, a custom linker \cite{LFZLS22} directly outlines machine code during the linking process without reliance on LTO, which greatly improved both code size and build time.
However, the potential of other linker optimizations is limited, as it involves elevating machine code to a higher-level IR, and the process of reconstructing metadata related to debug information and exception data is challenging.
Therefore, many other vital code size optimizations that leverage comprehensive program analysis are only achievable through LTO.

\paragraph*{Function merger.}

Function merging reduces code size by combining identical or similar functions into a single function.
LLVM offers a merge function pass \cite{mergefunc} at the IR level, but its efficiency can be affected by the linker's ICF, which performs a similar task. Previous attempts have been made to merge arbitrary pairs of functions \cite{ssafuncmerge,hyfm}. These approaches operate at the IR level, which simplifies transformations without machine-specific code generation concerns.

However, creating a standard function call can introduce higher call overhead, and precise cost modeling becomes more challenging due to interactions with subsequent passes, especially the function outliner.
Compared to the function outliner, the function merger targets entire functions, resulting in relatively limited merging opportunities and modest size impact, particularly when used with the function outliner and the linker's ICF.

Notably, none of the previous approaches, to our knowledge, support a scalable solution with separate compilations, a crucial requirement for building large-scale production binaries without incurring high compilation costs.
Addressing this limitation is a primary objective we aim to resolve in this paper.

\subsection{Global Function Outliner}\label{sec:globaloutliner}

\begin{figure}[!tb]
	\centering
	\begin{tikzpicture}

		\tikzstyle{profile}=[tape,tape bend top=none,draw,font=\scriptsize, fill=white, inner sep=1pt,minimum width=2.1cm, minimum height=1.1cm]
		\tikzstyle{profilecg}=[tape,tape bend top=none,draw,font=\scriptsize, fill=white, inner sep=1pt,minimum width=1.3cm, minimum height=1.1cm]
		\tikzstyle{compilation_step}=[draw,rectangle,minimum width=0.4cm, minimum height=0.9cm,align=center,font=\scriptsize]
		\tikzstyle{compilation_stage}=[thick, dashed, rectangle, draw,inner sep=3pt,font=\scriptsize]
    	\tikzstyle{pass}=[draw, rectangle, dashed, font=\scriptsize,minimum width=0.2cm,minimum height=0.78cm]

		\node[compilation_step] (thinlink) at (0,0) {
		 Thin\\
		 Link
		};
\begin{scope}[shift={(0.23,0)}]
		\node[compilation_step] (opt) at (.9,0) {Opt};
		\node[compilation_step] (cg1) at (1.9, 0) {CodeGen};

		\node[profile, fit=(opt)(cg1)](prf1) at (1.49, 0.2){};
		\node[profile, fit=(opt)(cg1)](prf1) at (1.52, 0.1){};
		\node[profile, fit=(opt)(cg1)](prf1) at (1.55, 0.0){};

		\node[compilation_step] (opt) at (0.94,0) {Opt};
		\node[compilation_step] (cg1) at (1.96, 0) {CodeGen};
		\node[pass](out1) at (2.36,-0.01){};
\end{scope}

		\node[compilation_step, rounded corners] (global) at (3.5,0) {
		 Global\\
		 Prefix\\
		 Tree
		};

\begin{scope}[shift={(0.45,0)}]
		\node[compilation_step] (cg2) at (4.5,0) {CodeGen};
	    \node[profilecg, fit=(cg2)](prf2) at (4.43, 0.2){};
	    \node[profilecg, fit=(cg2)](prf2) at (4.46, 0.1){};
	    \node[profilecg, fit=(cg2)](prf2) at (4.49, 0.0){};
		\node[compilation_step] (cg2) at (4.5,0) {CodeGen};
		\node[pass](out2) at (4.9,-0.01){};
\end{scope}

		\node[compilation_step] (link) at (6.4,0) {Linking};

		\node[rounded corners, draw, rectangle, minimum height=0.5cm, minimum width=5cm, text width=3cm,text centered,font=\scriptsize] (go) at (3.5, -1.5) {Global Function Outliner};
		\draw[->,very thick] (out1) edge node[left,font=\tiny] {analyze} (out1 |- go.north);

		\draw[->,very thick] (out2) edge node[left,font=\tiny] {outline} (out2 |- go.north);
		\draw[->,very thick] (go) edge node[left,font=\tiny] {combine} (go |- global.south);

	\end{tikzpicture}
	\caption{Overview of Global Function Outliner Pass.}
	\label{fig:globaloutline_overview}
\end{figure}
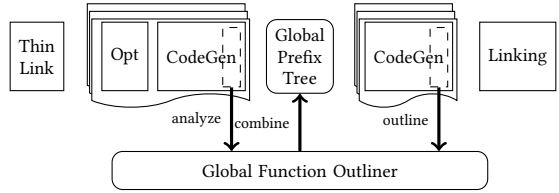

\begin{figure}[!tb]
	\centering
	\begin{tikzpicture}
	\tikzstyle{code}=[rounded corners,draw,font=\tiny, align=left,fill=white,minimum width=1cm,minimum height=0.5cm]
    \tikzstyle{module}=[thick, dashed, rectangle, draw,inner sep=3pt]

\begin{scope}[shift={(0,0)}]
    \node [code] (code1) at (0,0) {
      \textbf{\textcolor{blue}{f1:}}\\
      \hspace{0.1cm}mov x0, x1\\
      \hspace{0.1cm}lsl x0 ,x0 ,\#7\\
      \hspace{0.1cm}add x0 ,x8 ,\#77\\
      \hspace{0.1cm}ret
    };
    \node [code] (code2) at (0,-1.4) {
      \textbf{\textcolor{blue}{f2:}}\\
      \hspace{0.1cm}mov x0, x2\\
      \hspace{0.1cm}lsl x0 ,x0 ,\#7\\
      \hspace{0.1cm}add x0 ,x8 ,\#77\\
      \hspace{0.1cm}ret
    };

    \node [module, fit=(code1) (code2),label=above:{\scriptsize \bf Module1}] {};

    \node [code] (code6) at (0,-3.4) {
      \textbf{\textcolor{blue}{f3:}}\\
      \hspace{0.1cm}mov x0, x3\\
      \hspace{0.1cm}\textcolor{teal}{lsl x0,x0,\#7  \ \ \ //\textbf{Y}}\\
      \hspace{0.1cm}\textcolor{teal}{add x0,x8,\#77//\textbf{L}}\\
      \hspace{0.1cm}ret
    };
    \node [module, fit=(code6),label=above:{\scriptsize \bf Module2}] {};
\end{scope}

\begin{scope}[shift={(2.5,-2.8)}]
    \node[draw, fill=blue!10, ellipse, font=\tiny, label=above:{\tiny \bf GlobalPrefixTree}] (oval) at (0.08,0) {Root};
    \node[draw, font=\scriptsize] (y) at (0.08, -.7) {Y};
    \node[draw, very thick, font=\scriptsize] (L) at (-0.6, -1.2) {L};
    \node[draw, dotted, font=\scriptsize] (U) at (0.7, -1.2) {U};
    \draw[thick, ->] (oval) -- (y);
    \draw[thick, ->] (y) -- (L);
    \draw[dashed, thick, ->] (y) -- (U);
\end{scope}

\begin{scope}[shift={(5,.5)}]
    \node [code] (code3) at (0,0) {
      \textbf{\textcolor{blue}{f1:}}\\
      \hspace{0.1cm}mov x0, x1\\
      \hspace{0.1cm}\textcolor{black}{b outlinedFunc1}
    };
    \node [code] (code4) at (0,-.95) {
      \textbf{\textcolor{blue}{f2:}}\\
      \hspace{0.1cm}mov x0, x2\\
      \hspace{0.1cm}\textcolor{black}{b outlinedFunc1}
    };
    \node [code] (code5) at (0,-2.05) {
      \textbf{\textcolor{blue}{outlinedFunc1:}}\\
      \hspace{0.1cm}\textcolor{teal}{lsl x0,x0,\#7  \ \ \ //\textbf{Y}}\\
      \hspace{0.1cm}\textcolor{teal}{add x0,x8,\#77//\textbf{L}}\\
      \hspace{0.1cm}\textcolor{black}{ret}
    };
    \node [module, fit=(code3) (code5),label=above:{\scriptsize \bf Module1}] {};

    \node [code] (code7) at (0,-3.7) {
      \textbf{\textcolor{blue}{f3:}}\\
      \hspace{0.1cm}mov x0, x3\\
      \hspace{0.1cm}\textcolor{black}{b outlinedFunc2}
    };
    \node [code] (code8) at (0,-4.8) {
      \textbf{\textcolor{blue}{outlinedFunc2:}}\\
      \hspace{0.1cm}lsl x0 ,x0 ,\#7\\
      \hspace{0.1cm}add x0 ,x8 ,\#77\\
      \hspace{0.1cm}ret
    };
    \node [module, fit=(code7) (code8),label=above:{\scriptsize \bf Module2}] {};
\end{scope}

     \node [single arrow, fill=teal!10, draw, minimum width=1.2cm,font=\scriptsize, minimum height=0.5cm] (arrow) at (2.3,-1.3) {outline};
     \node [double arrow, fill=red!10, draw, minimum width=0.1cm,font=\tiny, minimum height=0.1cm] (arrow2) at (1.5,-3.2) {match};

    \draw[->, thick] (code3) to[out=195, in=165, looseness=0.6] (code5);
    \draw[->, thick] (code4) to[out=-15, in=15, looseness=1] (code5);
    \draw[->, thick] (code7) to[out=195, in=165, looseness=1] (code8);

\begin{scope}[shift={(9,-.4)}]
    \node [code] (code9) at (0,0) {
      \textbf{\textcolor{blue}{f1:}}\\
      \hspace{0.1cm}mov x0, x1\\
      \hspace{0.1cm}\textcolor{black}{b outlinedFunc1}
    };
    \node [code] (code10) at (0,-.95) {
      \textbf{\textcolor{blue}{f2:}}\\
      \hspace{0.1cm}mov x0, x2\\
      \hspace{0.1cm}\textcolor{black}{b outlinedFunc1}
    };
    \node [code] (code11) at (0,-1.9) {
      \textbf{\textcolor{blue}{f3:}}\\
      \hspace{0.1cm}mov x0, x3\\
      \hspace{0.1cm}\textcolor{black}{b outlinedFunc2}
    };
    \node [code] (code12) at (0,-2.95) {
      \textbf{\textcolor{blue}{outlinedFunc1/2:}}\\
      \hspace{0.1cm}{lsl x0,x0,\#7} \\
      \hspace{0.1cm}{add x0,x8,\#77}\\
      \hspace{0.1cm}\textcolor{black}{ret}
    };
    \node [module, fit=(code9) (code12),label=above:{\scriptsize \bf LinkedModule}] {};

    \draw[->, thick] (code9) to[out=195, in=165, looseness=0.5] (code12);
    \draw[->, thick] (code10) to[out=-15, in=15, looseness=0.6] (code12);
    \draw[->, thick] (code11) to[out=195, in=165, looseness=1] (code12);

\end{scope}

     \node [single arrow, fill=teal!10, draw, minimum width=1.2cm,font=\scriptsize, minimum height=0.5cm] (arrow) at (6.75,-2.0) {link};

	\end{tikzpicture}
	\caption{Global Function Outlining. The \GlobalPrefixTree was formed during the first round of \CodeGen from a local outlining instance, $outlinedFunc1$ using stable instruction hash sequences. In the second round of \CodeGen, $outlinedFunc2$ is optimistically outlined from $f3$ by finding the same hash sequence in the \GlobalPrefixTree. The linker can fold $outlinedFunc1$ and $outlinedFunc2$ since they are identical.}
	\label{fig:globaloutline_example}
\end{figure}
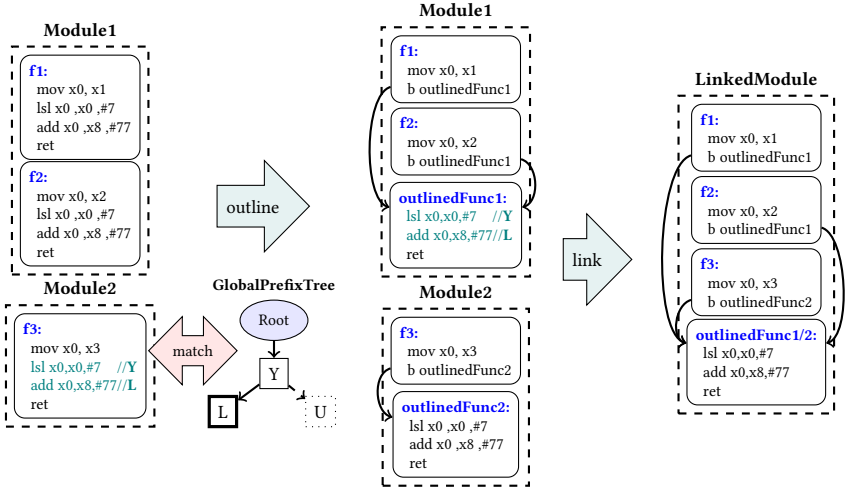

\gfo \cite{pgo-mobile} extends the scope of the function outliner \cite{machineoutliner} to a global level, particularly when used with ThinLTO.
Figure \ref{fig:globaloutline_overview} provides an overview of the ThinLTO pipeline.
After sequential thin-linking with bitcode summaries, each module undergoes parallel compilation, starting with optimization (\Opt) and code generation (\CodeGen).

In the first round of \CodeGen, the analyze step involves the function outliner operating locally, as usual, capturing outlined instruction sequences using stable hash sequences.
These hash sequences are efficiently published to a global prefix tree (\GlobalPrefixTree) during the subsequent combine step.
During the second round of \CodeGen, after local outlining of functions, the remaining instructions are further compared against the \GlobalPrefixTree to optimistically create an outlined function with a unique name.

Figure \ref{fig:globaloutline_example} provides an example.
In the first round of \CodeGen, $outlinedFunc1$ is created from two identical instruction sequences, namely, $lsl$ and $add$, locally within $Module1$.
The corresponding hashes, denoted as $Y$ and $L$, are added to the \GlobalPrefixTree.
In the second round of \CodeGen, although there is no explicit repetition in $Module2$, the instruction sequence in $f3$ matches with the \GlobalPrefixTree, leading $Module2$ to create an optimistic outlined function, $outlinedFunc2$.
At link-time, when both $outlinedFunc1$ and $outlinedFunc2$ are indeed identical,
the linker merges them via ICF.
The central idea behind this algorithm is that locally outlined instances might also appear in other modules.

However, we emphasize that outlined instances are not directly shared across independently operating modules.
Instead, we optimistically generate unique outlined functions that can be folded by the linker.
This principle is crucial to ensure soundness in the transformation.
Our paper is motivated by this approach, focusing on the optimistic creation of identical function merging instances from initially distinct yet similar functions.

\subsection{A Motivating Example}\label{sec:example}

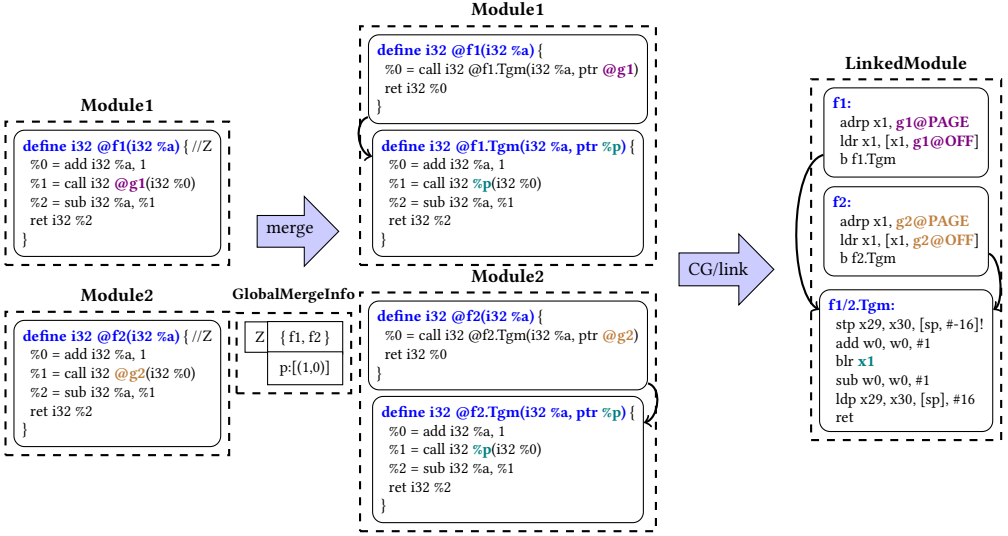
\begin{figure}[!tb]
	\centering
	\begin{tikzpicture}
	\tikzstyle{code}=[rounded corners,draw,font=\tiny, align=left,fill=white,minimum width=1cm,minimum height=0.5cm]
    \tikzstyle{module}=[thick, dashed, rectangle, draw,inner sep=3pt]
\begin{scope}[shift={(0,-1.5)}]
    \node [code] (code1) at (0,-.5) {
      \textbf{\textcolor{blue}{define i32 @f1(i32 \%a)}} \{\ //Z\\
      \hspace{0.1cm}\%0 = add i32 \%a, 1\\
      \hspace{0.1cm}{\%1 = call i32 \textbf{\textcolor{violet}{@g1}}(i32 \%0)}\\
      \hspace{0.1cm}{\%2 = sub i32 \%a, \%1}\\
      \hspace{0.1cm}{ret i32 \%2} \\
      \}
    };
    \node [module, fit=(code1),label=above:{\scriptsize \bf Module1}] {};

    \node [code] (code2) at (0,-3.0) {
      \textbf{\textcolor{blue}{define i32 @f2(i32 \%a)}} \{\ //Z\\
      \hspace{0.1cm}\%0 = add i32 \%a, 1\\
      \hspace{0.1cm}{\%1 = call i32 \textbf{\textcolor{brown}{@g2}}(i32 \%0)}\\
      \hspace{0.1cm}{\%2 = sub i32 \%a, \%1}\\
      \hspace{0.1cm}{ret i32 \%2} \\
      \}
      };
    \node [module, fit=(code2),label=above:{\scriptsize \bf Module2}] {};
\end{scope}

	\tikzstyle{hash}=[draw,font=\tiny,fill=white,minimum width=0.4cm,minimum height=0.4cm]
	\tikzstyle{val}=[draw,font=\tiny,fill=white,minimum width=1cm,minimum height=0.4cm]
\begin{scope}[shift={(1.9,-3.9)}]
    \node[hash] (h) at (0, +.01) {Z};
    \node[val] (sfs) at (0.6, 0) {\{ f1,\  f2 \}};
    \node[val] (param) at (0.6, -0.4) {p:[(1,0)]};
    \node [module, fit=(h) (sfs) (param),label=above:{\tiny \bf GlobalMergeInfo}] {};
\end{scope}

\begin{scope}[shift={(5.2,0)}]
    \node [code] (code3) at (0,-.5) {
      \textbf{\textcolor{blue}{define i32 @f1(i32 \%a)}} \{\\
      \hspace{0.1cm}\%0 = call i32 @f1.Tgm(i32 \%a, ptr \textbf{\textcolor{violet}{@g1}})\\
      \hspace{0.1cm}{ret i32 \%0} \\
      \}
    };
    \node [code] (code4) at (0,-2.0) {
      \textbf{\textcolor{blue}{define i32 @f1.Tgm(i32 \%a, ptr \textcolor{teal}{\%p})}} \{\\
      \hspace{0.1cm}{\%0 = add i32 \%a, 1}\\
      \hspace{0.1cm}{\%1 = call i32 \textbf{\textcolor{teal}{\%p}}(i32 \%0)}\\
      \hspace{0.1cm}{\%2 = sub i32 \%a, \%1}\\
      \hspace{0.1cm}{ret i32 \%2} \\
      \}
    };
    \node [module, fit=(code3) (code4),label=above:{\scriptsize \bf Module1}] {};

    \node [code] (code5) at (0,-4.0) {
      \textbf{\textcolor{blue}{define i32 @f2(i32 \%a)}} \{\\
      \hspace{0.1cm}\%0 = call i32 @f2.Tgm(i32 \%a, ptr \textbf{\textcolor{brown}{@g2}})\\
      \hspace{0.1cm}{ret i32 \%0} \\
      \}
    };
    \node [code] (code6) at (0,-5.5) {
      \textbf{\textcolor{blue}{define i32 @f2.Tgm(i32 \%a, ptr \textcolor{teal}{\%p})}} \{\\
      \hspace{0.1cm}{\%0 = add i32 \%a, 1}\\
      \hspace{0.1cm}{\%1 = call i32 \textbf{\textcolor{teal}{\%p}}(i32 \%0)}\\
      \hspace{0.1cm}{\%2 = sub i32 \%a, \%1}\\
      \hspace{0.1cm}{ret i32 \%2} \\
      \}
    };
    \node [module, fit=(code5) (code6),label=above:{\scriptsize \bf Module2}] {};
\end{scope}

    \draw[->, thick] (code3) to[out=195, in=165, looseness=1] (code4);
    \draw[->, thick] (code5) to[out=-15, in=15, looseness=1] (code6);

     \node [single arrow, fill=blue!20, draw, minimum width=0.5cm, minimum height=0.5cm] (arrow) at (2.3,-2.5) {\scriptsize merge};

\begin{scope}[shift={(10.5,-1.2)}]

    \node [code] (code7) at (0,0) {
      \textbf{\textcolor{blue}{f1:}}\\
      \hspace{0.1cm}{adrp x1, \textbf{\textcolor{violet}{g1@PAGE}}} \\
      \hspace{0.1cm}{ldr x1, [x1, \textbf{\textcolor{violet}{g1@OFF}}]} \\
      \hspace{0.1cm}\textcolor{black}{b f1.Tgm}
    };

    \node [code] (code8) at (0,-1.3) {
      \textbf{\textcolor{blue}{f2:}}\\
      \hspace{0.1cm}{adrp x1, \textbf{\textcolor{brown}{g2@PAGE}}} \\
      \hspace{0.1cm}{ldr x1, [x1, \textbf{\textcolor{brown}{g2@OFF}}]} \\
      \hspace{0.1cm}\textcolor{black}{b f2.Tgm}
    };
    \node [code] (code9) at (0,-3) {
      \textbf{\textcolor{blue}{f1/2.Tgm:}}\\
      \hspace{0.1cm}{stp	x29, x30, [sp, \#-16]!} \\
      \hspace{0.1cm}{add	w0, w0, \#1} \\
      \hspace{0.1cm}{blr	\textbf{\textcolor{teal}{x1}}} \\
      \hspace{0.1cm}{sub	w0, w0, \#1} \\
      \hspace{0.1cm}{ldp	x29, x30, [sp], \#16} \\
      \hspace{0.1cm}{ret}
    };
    \node [module, fit=(code7) (code9),label=above:{\scriptsize \bf LinkedModule}] {};
\end{scope}
    \draw[->, thick] (code7) to[out=195, in=150, looseness=0.6] (code9);
    \draw[->, thick] (code8) to[out=-15, in=30, looseness=0.6] (code9);

     \node [single arrow, fill=blue!20, draw, minimum width=0.5cm, minimum height=0.5cm] (arrow) at (8.0,-3.0) {\scriptsize CG/link};

	\end{tikzpicture}
	\caption{Global Function Merging. Using the global merge info, $f1$ and $f2$ optimistically create merging instances, $f1.Tgm$ and $f2.Tgm$, while supplying their own contexts, $@g1$ and $@g2$, respectively, via an additional parameter to the original constant location. $f1.Tgm$ and $f2.Tgm$ can be folded by the linker's ICF. }
	\label{fig:globalmerge_example}
\end{figure}

Initially, we were motivated by Swift's legacy merge function \cite{swiftmerge}, which effectively combines similar functions differing in constant operands.
This approach stands out from previous function merging techniques \cite{simfunc,ssafuncmerge,SeqAlign2019,hyfm} by preserving the integrity of function bodies while only replacing constants with parameters, without introducing extra control flows.
This preservation is crucial to retain other metadata, including debug information and critical transformations.
Despite its apparent simplicity, this approach is applicable to various instances in Swift's generated code, such as lazy protocol witness table accessors, protocol witnesses, getters, and more.
Additionally, we observed the applicability of this technique to other languages, including C++ templates and Objective-C blocks.
Thus we integrated this technique into the primary LLVM pipeline to make it accessible to a broader range of languages.
However, like other existing merge function techniques, this technique is also limited to local function merging within a single module, requiring direct IR comparisons.

Figure \ref{fig:globalmerge_example} is a motivating example illustrating the process of merging functions across two different modules that differ only by certain constants.
Specifically, the function $f1$ in $Module1$ and the function $f2$ in $Module2$ are almost identical except the call target constants, $g1$ and $g2$, respectively.
Our objective is to compute a function hash, denoted as $Z$, which represents the similarity between $f1$ and $f2$
while also maintaining the ability to track the locations of the differing constants.
This summary information is incorporated into the \GlobalMergeInfo, which will be elaborated upon in Section \ref{sec:algorithm}.

A key transformation creates parameterized merging instances, $f1.Tgm$ and $f2.Tgm$, independently for each module using \GlobalMergeInfo.
The original functions become thunks that supply their correct constant contexts to those merging instances which are optimistically identical. 
Subsequently, these two functions are separately lowered into machine code through \CodeGen.
If they remain identical, they can be efficiently folded by the linker's ICF.

\section{Algorithm}\label{sec:algorithm}
\subsection{Overview}\label{sec:overview}

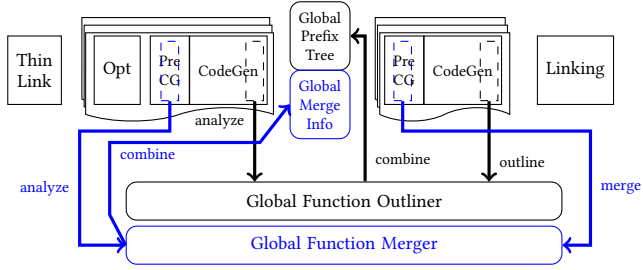
\begin{figure}[!tb]
	\centering
	\begin{tikzpicture}

		\tikzstyle{profile2}=[tape,tape bend top=none,draw,font=\scriptsize, fill=white, inner sep=1pt,minimum width=2.5cm, minimum height=1.1cm]
		\tikzstyle{profilecg2}=[tape,tape bend top=none,draw,font=\scriptsize, fill=white, inner sep=1pt,minimum width=1.73cm, minimum height=1.1cm]
		\tikzstyle{compilation_step}=[draw,rectangle,minimum width=0.4cm, minimum height=0.9cm,align=center,font=\scriptsize]
		\tikzstyle{compilation_stage}=[thick, dashed, rectangle, draw,inner sep=3pt,font=\scriptsize]
    	\tikzstyle{pass}=[draw, rectangle, dashed, font=\scriptsize,minimum width=0.2cm,minimum height=0.78cm]

		\node[compilation_step] (thinlink) at (0,0) {
		 Thin\\
		 Link
		};

\begin{scope}[shift={(0.2,0)}]
		\newcommand{\optcg}{
		\node[compilation_step] (opt) at (.9,0) {Opt};
		\node[compilation_step,font=\tiny] (pre1) at (1.6, 0) {Pre\\
CG
};
		\node[compilation_step,font=\tiny] (cg1) at (2.38, 0) {CodeGen};
		\node[pass,blue](out1m) at (1.6,-0.01){};
		\node[pass](out1) at (2.73,-0.01){};
		}

		\optcg
		\node[profile2, fit=(opt)(cg1)](prf1) at (1.68, 0.2){};
		\node[profile2, fit=(opt)(cg1)](prf1) at (1.71, 0.1){};
		\node[profile2, fit=(opt)(cg1)](prf1) at (1.73, 0.0){};
		\optcg

		\node[compilation_step,rounded corners,font=\tiny] (gpt) at (3.6,.45) {
		 Global\\
		 Prefix\\
		 Tree
		};
		\node[compilation_step, blue, rounded corners, font=\tiny] (gmi) at (3.6,.-.45) {
		 Global \\
		 Merge\\
		 Info
		};
\end{scope}

\begin{scope}[shift={(0.3,0)}]
		\newcommand{\optcgp}{
		\node[pass,blue](out2m) at (4.6,-0.01){};
		\node[compilation_step, font=\tiny] (pre2) at (4.62, 0) {Pre\\
CG
};
		\node[compilation_step, font=\tiny] (cg2) at (5.4,0) {CodeGen};
	    \node[pass](out2) at (5.75,-0.01){};
		}

	    \optcgp
 	    \node[profilecg2, fit=(pre2)(cg2)](prf2) at (5.10, 0.2){};
 	    \node[profilecg2, fit=(pre2)(cg2)](prf2) at (5.13, 0.1){};
 	    \node[profilecg2, fit=(pre2)(cg2)](prf2) at (5.16, 0.0){};
	    \optcgp
\end{scope}

		\node[compilation_step] (link) at (7.2,0) {Linking};

		\node[rounded corners, draw, rectangle, minimum height=0.5cm, minimum width=5.8cm, text width=3cm,text centered,font=\scriptsize] (go2) at (4.12, -1.72) {Global Function Outliner};
		\node[rounded corners, draw, rectangle, minimum height=0.5cm, minimum width=5.8cm, text width=3cm,text centered,font=\scriptsize, blue] (gm2) at (4.12, -2.3) {Global Function Merger};

		\draw[->,very thick] (out1) edge node[near start, left,font=\tiny] {analyze} (out1 |- go2.north);
		\draw[->,very thick] (out2) edge node[right,near end, font=\tiny] {outline} (out2 |- go2.north);
        \draw[->,very thick] (4.4, -1.46)
            to [out=90, in=270, looseness=0] node[near start, right, font=\tiny] {combine} (4.4, 0.45)
            to [out=180, in=0, looseness=0] (gpt.east);

		\draw[->,very thick,blue] (out1m.south) to [out=-90, in=90, looseness=0] (1.8, -0.8)
		    to [out=-90, in=0, looseness=0] (.6, -0.8)
		    to [out=270, in=90, looseness=0] node[midway, left, font=\tiny] {analyze} (0.6, -2.3)
		    to (gm2.west);
		\draw[->,very thick,blue] (gm2.west) to [out=150, in=-30, looseness=0] (1.0,-1.9)
		    to [out=90, in=270, looseness=0] node[near end, right, font=\tiny] {combine} (1.0, -0.95)
		    to [out=0, in=180, looseness=0] (3., -0.95)
		    to [out=90, in=180, looseness=0] (gmi.west);
		\draw[->,very thick, blue] (out2m.south) to [out=-90, in=90, looseness=0] (4.9, -0.8)
		    to [out=90, in=0, looseness=0] node[near end, right, font=\tiny] {} (7.4, -0.8)
            to [out=-90, in=90, looseness=0] node[midway, right, font=\tiny] {merge} (7.4, -2.3)
            to (gm2.east);
	\end{tikzpicture}
	\begin{tikzpicture}
\end{tikzpicture}
	\caption{Overview of Global Function Merger with Global Function Outliner}
	\label{fig:globalmerger_overview}
\end{figure}

Figure \ref{fig:globalmerger_overview} shows an overview of \gfm, which builds upon \gfo shown in Figure \ref{fig:globaloutline_overview}.
Like \gfo, the process consists of three steps.

In the first step (Section \ref{sec:analyze}), during the first-round \CodeGen, functions within each module are individually analyzed to compute stable function summaries.
In the subsequent step (Section \ref{sec:combine}), these summaries are combined to produce \GlobalMergeInfo.
In the final step (Section \ref{sec:merge}), during the second-round \CodeGen, merged functions are optimistically created within each module using \GlobalMergeInfo.
Unlike \gfo, \gfm operates at a late IR pass, as discussed in Section \ref{sec:background}. These two optimizations are integrated into a common \CodeGen framework by pushing \gfm down to the pre-\CodeGen pass while operating on the IR as before.

As \gfo and \gfm are separate passes, they can potentially interact with each other when using global summaries from the previous \CodeGen steps independently.
Specifically, in the first-round \CodeGen, the \GlobalPrefixTree is based on functions that have not been merged yet.
In the second-round \CodeGen, once merging function instances are created by \gfm, the previously computed \GlobalPrefixTree does not reflect the changes in these functions.
Ideally, running the three \CodeGen passes in the following order would be optimal: (i) analyze merging opportunities to produce \GlobalMergeInfo, (ii) merge functions while analyzing outlining opportunities with \GlobalPrefixTree, and (iii) finally, both merge and outline functions using \GlobalMergeInfo and \GlobalPrefixTree.

However, we have observed that, unlike \GlobalMergeInfo, \GlobalPrefixTree tends to capture meaningful outlining opportunities even when there are changes in the IR, as \gfo typically targets short sequences of machine code that are commonly found.
Further discussion on this topic can be found in Section \ref{sec:interaction} with supporting data in Section \ref{sec:tradeoff}.
Therefore, repeating \CodeGen once while running \gfo and \gfm in order remains practical and effective.
Finally, in Section \ref{sec:soundness}, we delve into the soundness of our approach and its overall time complexity.

\newcommand{\call}{\textit{Call}\xspace}
\newcommand{\gep}{\textit{GetElementPtr}\xspace}
\newcommand{\load}{\textit{Load}\xspace}
\newcommand{\store}{\textit{Store}\xspace}
\newcommand{\invoke}{\textit{Invoke}\xspace}

\begin{algorithm2e}[!tb]
\DontPrintSemicolon
  \caption{Analyze Functions}
  \label{algo:analyze}
  \vspace{-0.3cm}
  \begin{multicols}{2}
  \SetKwInOut{Input}{Input}
  \SetKwInOut{Output}{Output}
  \SetKwProg{Fn}{Function}{}{end}
  \Input{$Module$ \cmt{Compilation Unit}}
  \Output{$StableFns$ \cmt{Stable Function Summaries}}

  \SetKwFunction{Analyze}{Analyze}
  \SetKwFunction{ComputeStableFn}{ComputeStableFn}
  \SetKwFunction{isValid}{isValid}
  \SetKwFunction{isEligible}{isEligible}
  \SetKwFunction{isParametrizable}{isParametrizable}
  \SetKwFunction{canParam}{canParam}
  \SetKwFunction{combineHash}{combineHash}
  \SetKwFunction{hash}{hash}
  \SetKwFunction{getOpcode}{getOpcode}
  \BlankLine

  \Fn(){\Analyze($Module$)}{
    \For{$function \in Module$} {
      $SF \leftarrow \ComputeStableFn(function)$\;
      \If{\isValid($SF$)} {
        $StableFns.add(SF)$\;
      }
    }
    \Return{$StableFns$}\;
  }

  \Fn(){\ComputeStableFn($function$)}{
      $i \leftarrow 0$ \cmt{Instruction index}
      $H \leftarrow 0$\;
      \For{$instruction \in function$} {
        $opc \leftarrow \getOpcode(instruction)$\;
        $H \leftarrow H \otimes \hash(opc)$\;
        $j \leftarrow 0$ \cmt{Operand index}
        \For{$opnd \in instruction$} {
          $H_{opnd} \leftarrow \hash(opnd)$\;
          \If{$\canParam(opc, opnd$)} {
            $locToHash[(i,j)] \leftarrow H_{opnd}$\;
          }
          \Else {
            $H \leftarrow H \otimes  H_{opnd}$\;

          }
          $j \leftarrow j + 1$\;
        }
        $i \leftarrow i + 1$\;
      }
    \Return{$(H, modName, fnName, i, locToHash)$}\;
  }
  \end{multicols}
\end{algorithm2e}

\subsection{Analyze}\label{sec:analyze}
\begin{wrapfigure}{r}{0.45\textwidth}
  \centering
    \includegraphics[width=0.45\textwidth]{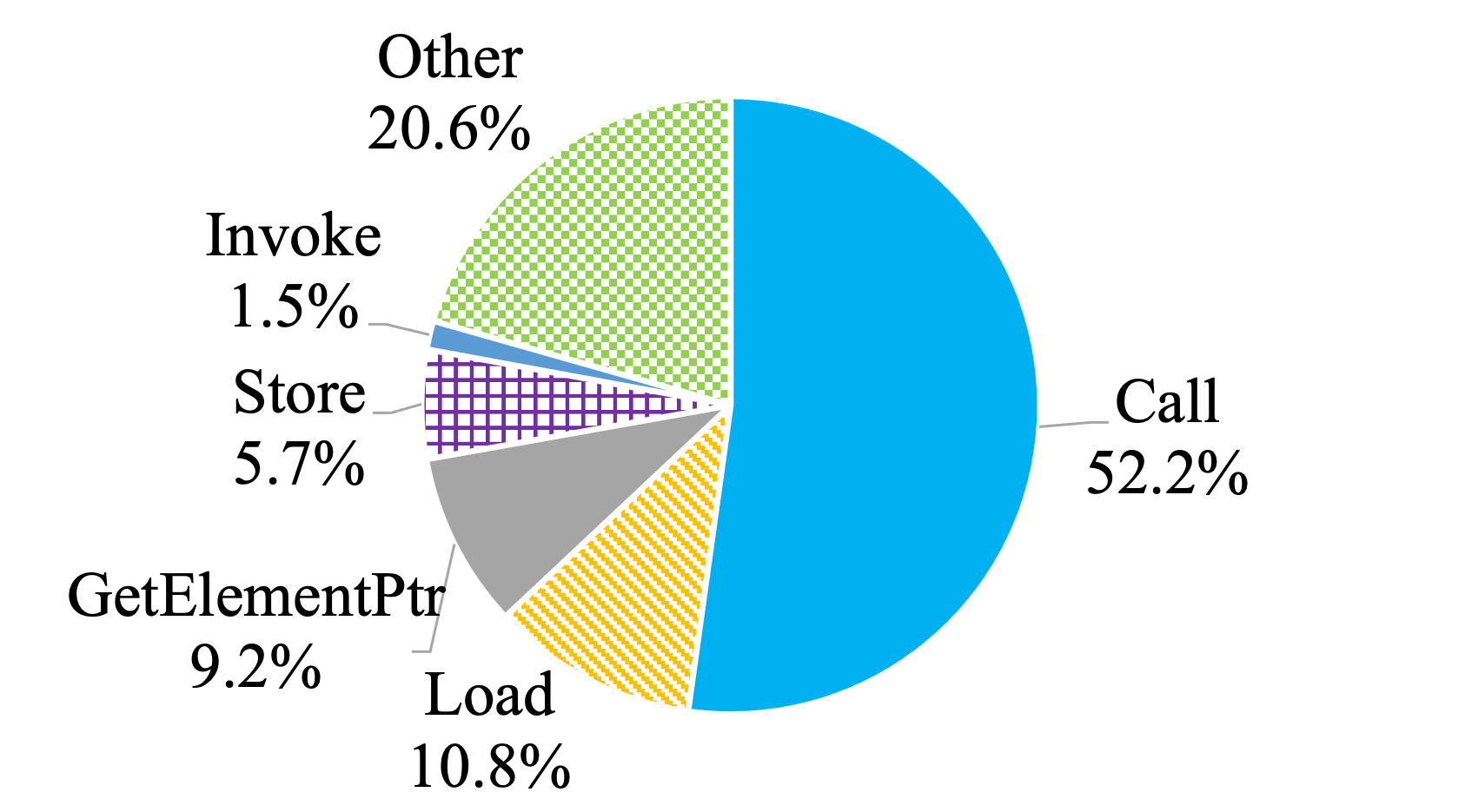}
    \caption{Percentage Share of Instruction Opcode Category for \fb.}
    \label{fig:instructionmix}
\end{wrapfigure}
Algorithm \ref{algo:analyze} outlines the process by which we independently analyze functions within a given module and generate a stable function summary that can be operated on later without the IR.
For each function, we iterate through its instructions, combining the hashes of opcodes and operands.
Our primary goal is to compute a combined hash while ignoring constant hashes for eligible instructions.
We maintain a record of the mapping between the ignored constant locations and and their respective hashes.
The locations are denoted as $(i, j)$ where $i$ is the index of the instruction and $j$ is the index of the operand.
In each stable function summary we publish the combined stable hash, along with the module name, function name, instruction count, and the mapping of locations to hashes that were skipped.

Figure \ref{fig:instructionmix} presents the percentage distribution of instruction opcode categories for \fb at the IR level.
Notably, more than 50\% of these instructions are \call instructions, where direct \call instructions target constant function symbols.
The remaining opcodes are \load and \store, which generally target constant global symbols.
\gep also accounts for a significant portion at 9\%.
However, \gep takes a type as the first operand, which can be challenging to canonicalize across modules where not all types are available.
\invoke shares a similar semantic with \call except that it deals with exceptions.
Thus, in \canParam(), we focus on tracking just four types of instructions: \call, \load, \store, and \invoke.
These instructions collectively cover more than 70\% of the instructions while their constant operands can be easily parameterized.






\subsection{Combine}\label{sec:combine}
All stable function summaries are generated during the previous \CodeGen pass that runs in parallel.
In this step, we sequentially combine these summaries.
As outlined by Algorithm \ref{algo:combine}, we detail the process of grouping stable functions and determining the set of parameters using the location-to-hash mapping.
Initially we use the combined hash to group stable functions that are merging candidates.
Then, we validate and filter out those merging candidates in the \textit{canMerge()} process by ensuring they are structurally identical by matching instruction count and the location-to-hash mapping.


\begin{algorithm2e}[!t]
\DontPrintSemicolon
  \caption{Combine Function Summaries}
  \label{algo:combine}
  \vspace{-0.3cm}
  \begin{multicols}{2}

  \SetKwInOut{Input}{Input}
  \SetKwInOut{Output}{Output}
  \SetKwProg{Fn}{Function}{}{end}
  \SetKwFunction{canMerge}{canMerge}
    \SetKwFunction{shouldMerge}{shouldMerge}
    \SetKwFunction{Combine}{Combine}
    \SetKwFunction{ComputeParams}{ComputeParams}
  \Input{$StableFns$ \cmt{Published from all modules}}
  \Output{$GlobalMergeInfo$}
  \BlankLine

  \Fn(){\Combine($StableFns$)}{
    \cmt{Group stable functions per stable hash.}
    \For{$SF \in StableFns$} {
      $HashToStableFns[SF.H].add(SF)$\;
    }

    \For{$(H, SFS) \in HashToStableFns$} {
      \cmt{Validate merging candidates.}
      \If{$!\canMerge(SFS)$} {
        \textbf{continue}\;
      }

      $ParamVecs \leftarrow \ComputeParams(SFS)$ \;

      \cmt{Apply a cost model from Equation \ref{eq:mergecost}.}
      \If{$\shouldMerge(SFS, ParamVecs)$} {
      $GlobalMergeInfo[H] \leftarrow (SFS, ParamVecs)$\;
      }
    }
    \Return{$GlobalMergeInfo$}\;
  }
  \cmt{Find distinct constant hash sequence.}
  \Fn(){\ComputeParams($SFS$)}{
      $RSF \leftarrow SFS[0]$\;
      \For{$(Loc, Hash) \in RSF.LocToHash$} {
        $CHashSeq \leftarrow [Hash]$\;
        $Distinct \leftarrow false$\;
        \For{$SF \in SFS[1:]$} {
          $SHash \leftarrow SF.LocToHash[Loc]$\;
          \If{$SHash \neq Hash$} {
             $Distinct \leftarrow true$\;
          }
          $CHashSeq.add(SHash)$\
        }
        \cmt{Append a location to the unique hash sequence.}
        \If{$Distinct$} {
        $ParamVecs[CHashSeq].add(Loc)$\
        }
    }
    \Return{$ParamVecs$}\;
  }
  \end{multicols}
\end{algorithm2e}

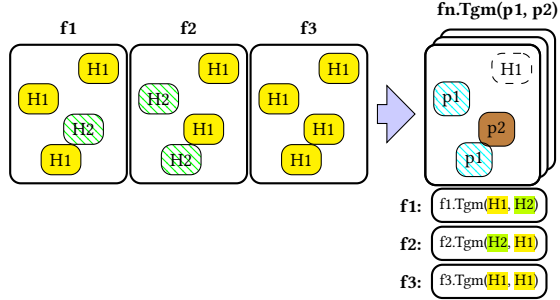
\begin{figure}[h]
	\begin{tikzpicture}
	\tikzstyle{code}=[rounded corners,draw,font=\scriptsize, align=left,fill=white,minimum width=1.7cm,minimum height=0.5cm]
    \tikzstyle{function}=[thick, rounded corners, rectangle, draw,inner sep=3pt]
	\tikzstyle{const1}=[rounded corners,draw,font=\scriptsize,fill=white,minimum width=0.08cm,minimum height=0.08cm]
\begin{scope}[shift={(0,-0.1)}]
\begin{scope}[shift={(0,0)}]
    \node [const1,fill=yellow] (c1) at (0.8,0,0) {H1};
    \node [const1,fill=yellow] (c2) at (0,-0.4) {H1};
    \node [const1, pattern=north west lines, pattern color=green] (c3) at (0.6,-0.8) {H2};

    \node [const1,fill=yellow] (c4) at (0.3,-1.2) {H1};
    \node [function, fit=(c1) (c2) (c3) (c4),label=above:{\scriptsize \bf f1}] {};
\end{scope}
\begin{scope}[shift={(1.6,0)}]
    \node [const1,fill=yellow] (c1) at (0.8,0,0) {H1};
    \node [const1,pattern=north west lines, pattern color=green] (c2) at (0,-0.4) {H2};
    \node [const1, fill=yellow] (c3) at (0.6,-0.8) {H1};
    \node [const1,pattern=north west lines, pattern color=green] (c4) at (0.3,-1.2) {H2};
    \node [function, fit=(c1) (c2) (c3) (c4),label=above:{\scriptsize \bf f2}] {};
\end{scope}
\begin{scope}[shift={(3.2,0)}]
    \node [const1,fill=yellow] (c1) at (0.8,0,0) {H1};
    \node [const1,fill=yellow] (c2) at (0,-0.4) {H1};
    \node [const1, fill=yellow] (c3) at (0.6,-0.8) {H1};
    \node [const1,fill=yellow] (c4) at (0.3,-1.2) {H1};
    \node [function, fit=(c1) (c2) (c3) (c4),label=above:{\scriptsize \bf f3}] {};
\end{scope}

\end{scope}

\newcommand\paramseq{
    \node [const1,dashed] (c1) at (0.8,0,0) {H1};
    \node [const1,pattern=north west lines, pattern color=cyan] (c2) at (0,-0.4) {p1};
    \node [const1, fill=brown] (c3) at (0.6,-0.8) {p2};
    \node [const1,pattern=north west lines, pattern color=cyan] (c4) at (0.3,-1.2) {p1};

}

\begin{scope}[shift={(5.7,0.1)}]
    \paramseq
    \node [function, fit=(c1) (c2) (c3) (c4), label=above:{\scriptsize \bf fn.Tgm(p1, p2)}, fill=white] {};
\end{scope}

\begin{scope}[shift={(5.6,0)}]
    \paramseq
    \node [function, fit=(c1) (c2) (c3) (c4),fill=white] {};
\end{scope}

\begin{scope}[shift={(5.5,-0.1)}]
    \paramseq
    \node [function, fit=(c1) (c2) (c3) (c4),fill=white] {};
\end{scope}
\begin{scope}[shift={(5.5,-0.1)}]
    \paramseq
    \node [function, fit=(c1) (c2) (c3) (c4)] {};
\end{scope}

     \node [single arrow, fill=blue!20, draw, minimum width=0.5cm, minimum height=0.5cm] (arrow) at (4.7,-0.7) {};

\begin{scope}[shift={(6,-1.9)}]
    \tikzstyle{function2}=[thick, font=\tiny, rounded corners, rectangle, draw,inner sep=3pt]
    \node [function2,label=left:{\scriptsize \bf f1:}] at (0,0) {
        f1.Tgm(\sethlcolor{yellow}\hl{H1}, \sethlcolor{lime}\hl{H2})
    };
    \node [function2,label=left:{\scriptsize \bf f2:}] at (0,-.5){
        f2.Tgm(\sethlcolor{lime}\hl{H2}, \sethlcolor{yellow}\hl{H1})
    };
    \node [function2,label=left:{\scriptsize \bf f3:}] at (0,-1) {
        f3.Tgm(\sethlcolor{yellow}\hl{H1}, \sethlcolor{yellow}\hl{H1})
    };
\end{scope}

	\end{tikzpicture}
	\caption{Assigning Parameters. Functions $f1$, $f2$ and $f3$ are assumed to have 4 constant operands at distinct locations and their hashes, $H1$ or $H2$ that are ignored, making their function hashes identical. Two parameters, $p1$ and $p2$ are allocated to supply different constant contexts to the identically parameterized functions.  }
	\label{fig:parameters}
\end{figure}

To assign the set of parameters, we focus on identifying unique sequences of constant hashes across stable functions within the same group.
Figure \ref{fig:parameters} provides an example of parameter assignments for functions $f1$, $f2$, and $f3$, each with four constant operands tracked in the location-to-hash mapping.
If all the constant hashes for a given location are identical across the grouped stable functions, no parameter allocation is necessary, as shown by the dashed box in the 1\textsuperscript{st} sequence: $(H1, H1, H1)$.
However, when constant hash sequences differ, we use a map to identify unique constant hash sequences and accumulate their corresponding locations.
For instance, the sequence $(H1, H2, H1)$ is repeated in the 2\textsuperscript{nd} and 4\textsuperscript{th} locations and can be assigned to a single parameter, denoted as $P1$. The remaining sequence, $(H2, H1, H1)$, is assigned to a different parameter, represented as $P2$.

 \begin{flalign} \label{eq:mergecost}
     \begin{aligned}
         &Benefit = Size_{func} * (N - 1) \\
         &Cost = Size_{thunk} * N \\
         &\text{\;where~$N$ is the number of merged functions,} \\
         &\quad \qquad \text{$Size_{func}$ is the function size, and } \\
         &\quad \qquad \text{$Size_{thunk}$ is the thunk size}  \\
         &\quad \qquad \text{including parameters, call, and metadata.}
     \end{aligned}
 \end{flalign}

Eventually these original functions will be transformed into thunks that call the commonly canonicalized functions identified as $f1.Tgm$, $f2.Tgm$ and $f3.Tgm$, respectively, and pass varying constants as parameters.
In \shouldMerge(), we employ Equation \ref{eq:mergecost} to calculate the potential reduction in size achieved by consolidating $N$ functions into a single one, along with the overall size cost associated with adding $N$ thunks.
If the condition $Cost < Benefit$ holds true, such stable functions are considered profitable and published to the \textit{GlobalMergeInfo}.

\subsection{Merge}\label{sec:merge}

  \SetKwInOut{Input}{Input}
  \SetKwInOut{Output}{Output}
  \SetKwProg{Fn}{Function}{}{end}
  \SetKwFunction{Merge}{Merge}
    \SetKwFunction{Match}{Match}
    \SetKwFunction{isCompatiable}{isCompatiable}
    \SetKwFunction{createMergedFunc}{createMergedFunc}
    \SetKwFunction{createThunk}{createThunk}
    \SetKwFunction{ComputeStableFn}{ComputeStableFn}
    \SetKwFunction{getArgs}{getArgs}

\begin{algorithm2e}[!t]
\DontPrintSemicolon
  \caption{Merge Functions}
  \label{algo:merge}
  \vspace{-0.3cm}

  \begin{multicols}{2}
  \Input{$GlobalMergeInfo$, $Module$}
  \Output{$Module$ \cmt{Optimized (merged) Module}}
  \BlankLine

  \Fn(){\Merge($GlobalMergeInfo$, $Module$)}{
    \For{$(H, SFS, ParamVecs) \in GlobalMergeInfo$} {
      $MergeInfo \leftarrow \Match(SFS, Module)$\;
      \For{$MI \in MergeInfo$} {
        \cmt{Get the original constants from locations in $MI.f$ pointed by $ParamVecs$.}
        $Args \leftarrow \getArgs(MI, ParamVecs)$\;
  \BlankLine
        \cmt{Create $MergedFunc$ with parameters.}
        $MergedFunc \leftarrow \createMergedFunc(MI, ParamVecs)$\;

        \cmt{The function becomes a thunk to call $MergedFunc$ with the original constants.}
        $\createThunk(MI, MergedFunc, Args)$\;
      }
    }
    \Return{$Module$}\;
  }
  \Fn(){\Match($SFS$, $Module$)}{
      $ModName \leftarrow Module.name$\;
      $Local \leftarrow true$\;
      $RModName \leftarrow SFS[0].modName$\;
      \For{$SF \in SFS$} {
         $SModName \leftarrow SF.modName$\;
        \If{$RModName \neq ModName$} {
            $Local \leftarrow false$\;
        }
        \If{$ModName \neq SModName$} {
            \textbf{continue}\;
        }

        $f \leftarrow Module.find(SF.fnName)$\;
        $SFC \leftarrow \ComputeStableFn(f)$\;
        \If{$\isCompatiable(SF, SFC)$} {
            $MergeInfo.add([SF, f])$\;
        }
      }
      \cmt{Skip if it's a single local candidate.}
      \If{$Local ~~\mathrm{ \textbf{and} }~~ len(MergeInfo) < 2$} {
        \Return \{\}\;
      } \Else {
      \Return $MergeInfo$\
      }
  }
  \end{multicols}
\end{algorithm2e}

The \textit{GlobalMergeInfo} has been computed in the preceding step.
In Figure \ref{fig:globalmerge_example}, you can see that two functions, $f1$ and $f2$, are grouped based on their function hash, represented as $Z$.
This hash calculation excludes constant operands ($g1$ and $g2$) found at the same location $(1,0)$, which corresponds to the instruction and operand index pair.
This information is captured in the \GlobalMergeInfo.

Algorithm \ref{algo:merge} outlines the process of gathering merge information by matching functions in the current module against stable function summaries.
In \Match(), we compute stable function summaries to evaluate the compatibility of functions with matching module and function names as merging candidates.
Even if the merging candidate is unique within the current module, we optimistically proceed to create a parameterized merging instance, unless there are no cross-module candidates in the stable function summary.

When creating a parameterized merging instance, we populate the actual constants located at \textit{ParamVecs} from the original function.
Afterward, we canonicalize the original function into a merged function by replacing these constant locations with parameters.
The merged function's name is suffixed with $.Tgm$ for distinction.
We then generate a thunk that preserves the original constants and passes them to the merged function.
During the linking process shown in Figure \ref{fig:globalmerge_example}, these merged functions are consolidated by the linker's ICF when they are indeed identical.

\subsection{Soundness and Complexity}\label{sec:soundness}
Unlike traditional function merging, which relies on precise comparisons of IR, our approach is primarily based on hash summaries.
This choice is driven by the unavailability of IR when using separate compilations, which we are targeting.
Since we cannot entirely eliminate the possibility of hash collisions or the potential staleness of these summaries, it is imperative that our transformation remains safe.
Since each module uses the \textit{GlobalMergeInfo} independently, forcefully transforming the IR to directly share the merged function instances or modify the call-sites can result in mis-compilation.

Instead, our approach focuses on canonicalizing each merging candidate individually, without creating a shared merging instance among them.
We conservatively assume original functions may be accessed across modules and consistently create a thunk, without any attempts to update the call-site in place.
This approach provides the flexibility to merge functions in any order, accommodating cases of recursive function merging.
This transformation guarantees safety, albeit at the cost of size efficiency.
It's only when the merged functions are genuinely identical at link time that they can be safely folded, resulting in significant savings in overall binary size.

Our algorithms exhibit nearly linear time complexity in all three cases.
In Algorithm \ref{algo:analyze}, we perform a single pass for each function to create stable function summaries.
In Algorithm \ref{algo:combine}, we group all stable function summaries based on their function hashes and allocate parameters among the grouped function summaries in a single pass.
In Algorithm \ref{algo:merge}, we identify functions that match the \textit{GlobalMergeInfo} summary and transform them into merged functions and thunks, all within a single pass.
However, it's worth noting that our transformation results in more identical functions, which may increase the time required for the identical code folding process during linking.

\section{Implementation}\label{sec:implementation}
We implemented two function mergers in Clang version 16.0 using the LLD linker.
As described in Section \ref{sec:example}, we initially ported Swift's legacy merge function \cite{swiftmerge} into the LTO pipeline as a local function merger (\lfm), which serves as our baseline implementation for comparison.
For \gfm, we implemented a structural hash function \cite{Grossman2023} following a structure similar to the function comparator in \lfm.
As depicted in Figure \ref{fig:globalmerger_overview}, \gfm is integrated during the pre-\CodeGen pass, allowing us to incorporate the common \CodeGen framework for both \gfm and \gfo.

We explored several implementation details and practical aspects in this section.
In Section \ref{sec:singlecg}, we experimented with an alternative approach, implementing \gfm using a single \CodeGen pass similar to NoLTO while utilizing \CodeGen artifacts published from previous builds.
This trade-off between build time and size optimization was part of our considerations.
In Section \ref{sec:interaction}, we discussed interactions between \gfo and \gfm, and made modifications to \gfo to deterministically outline the merging candidates produced by \gfm.
In Section \ref{sec:objcglobal}, we enhanced the precision of the hash function for Objective-C globals prevalent in mobile apps.
Finally, in Section \ref{sec:debug}, we discuss how our work can assist developers in triaging stack traces related to merged functions.

\subsection{Single Codegen Alternative}\label{sec:singlecg}
In Section \ref{sec:algorithm}, we explain that we run \CodeGen twice.
The initial run is used to analyze the IR or the MIR for potential merging or outlining opportunities.
Subsequently, we combine these results and leverage them during the second-round \CodeGen pass to generate more common code that can be folded by the linker.
While it may seem wasteful to run \CodeGen twice, it turns out that \CodeGen makes up a relatively small portion of the overall build time.
Nevertheless, it is important to reduce build time as much as possible, particularly, in production environments where it helps to save build resources for other essential workloads.

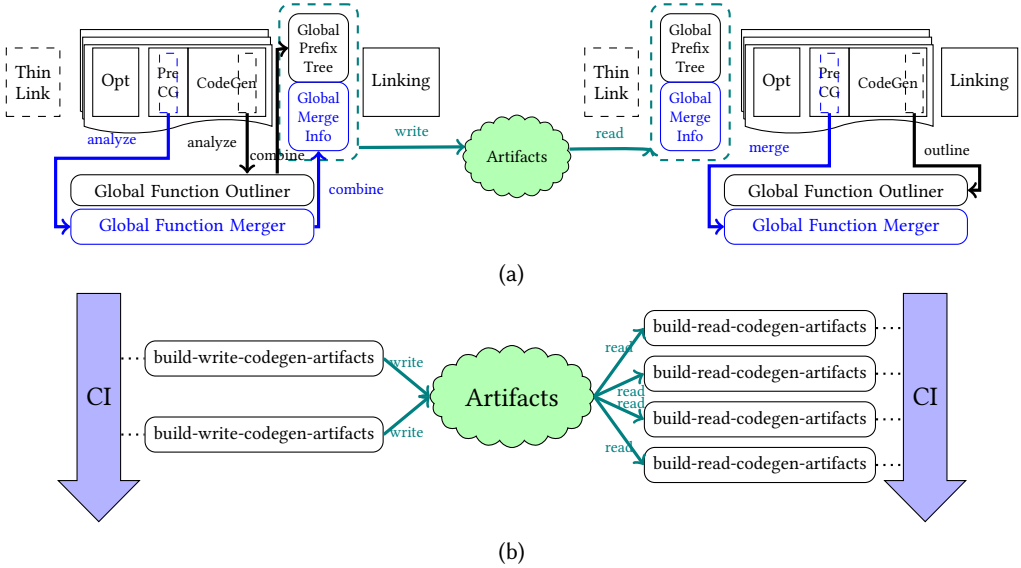
\begin{figure}[!tb]

\begin{subfigure}[t]{\columnwidth}
	\centering
	\begin{tikzpicture}

	\begin{scope}[shift={(0,0)}]
		\tikzstyle{profile2}=[tape,tape bend top=none,draw,font=\scriptsize, fill=white, inner sep=1pt,minimum width=2.5cm, minimum height=1.1cm]
		\tikzstyle{profilecg2}=[tape,tape bend top=none,draw,font=\scriptsize, fill=white, inner sep=1pt,minimum width=1.73cm, minimum height=1.1cm]
		\tikzstyle{compilation_step}=[draw,rectangle,minimum width=0.4cm, minimum height=0.9cm,align=center,font=\scriptsize]
		\tikzstyle{compilation_stage}=[thick, dashed, rectangle, draw,inner sep=3pt,font=\scriptsize]
    	\tikzstyle{pass}=[draw, rectangle, dashed, font=\scriptsize,minimum width=0.2cm,minimum height=0.78cm]

		\node[compilation_step,dashed] (thinlink) at (1.1,0) {
		 Thin\\
		 Link
		};

	\begin{scope}[shift={(1.3,0)}]
		\newcommand{\optcg}{
		\node[compilation_step] (opt) at (.9,0) {Opt};
		\node[compilation_step,font=\tiny] (pre1) at (1.6, 0) {Pre\\
CG
};
		\node[compilation_step,font=\tiny] (cg1) at (2.38, 0) {CodeGen};
		\node[pass,blue](out1m) at (1.6,-0.01){};
		\node[pass](out1) at (2.65,-0.01){};
		}

		\optcg
		\node[profile2, fit=(opt)(cg1)](prf1) at (1.68, 0.2){};
		\node[profile2, fit=(opt)(cg1)](prf1) at (1.71, 0.1){};
		\node[profile2, fit=(opt)(cg1)](prf1) at (1.73, 0.0){};
		\optcg
	\end{scope}

		\node[compilation_step,rounded corners,font=\tiny] (gpt) at (4.9,.45) {
		 Global\\
		 Prefix\\
		 Tree
		};
		\node[compilation_step, blue, rounded corners, font=\tiny] (gmi) at (4.9,.-.45) {
		 Global \\
		 Merge\\
		 Info
		};
          \node[compilation_step,rounded corners,teal,thick,dashed,fit=(gpt)(gmi)] (write) at (4.9,0){};

		\node[compilation_step] (link) at (6,0) {Linking};

		\node[rounded corners, draw, rectangle, minimum height=0.3cm, minimum width=2.5cm, text width=3cm,text centered,font=\scriptsize] (go2) at (3.22, -1.42) {Global Function Outliner};
		\node[rounded corners, draw, rectangle, minimum height=0.3cm, minimum width=2.5cm, text width=3cm,text centered,font=\scriptsize, blue] (gm2) at (3.22, -1.9) {Global Function Merger};

		\draw[->,very thick,blue](gm2.east)
		    to [out=90, in=180, looseness=0]  (4.9, -1.9)
		    to node[right, font=\tiny] {combine} (gmi.south);
		\draw[->,very thick](4.35, -1.2)
		    to [out=270, in=0, looseness=0] node[near start, font=\tiny] {combine} (4.35, 0.43)
		    to (gpt.west);

	    \draw[->,very thick, teal](write)
		    to [out=302, in=0, looseness=0] node[font=\tiny, above] {write} (6.85,-.85);

		\draw[->,very thick,blue] (out1m.south)
		    to [out=-90, in=180, looseness=0] (2.9, -1.0)
		    to [out=0, in=-90, looseness=0] node[midway, above, font=\tiny] {analyze} (1.4, -1.0)
		    to [out=270, in=90, looseness=0] (1.4, -1.9)
		    to (gm2.west);

		\draw[->,very thick] (out1.south)
		    to [out=90, in=90, looseness=0] node[midway, left, font=\tiny] {analyze}(3.95, -1.2);
\end{scope}

\begin{scope}[shift={(7.7,0)}]
		\tikzstyle{profile2}=[tape,tape bend top=none,draw,font=\scriptsize, fill=white, inner sep=1pt,minimum width=2.5cm, minimum height=1.1cm]
		\tikzstyle{profilecg2}=[tape,tape bend top=none,draw,font=\scriptsize, fill=white, inner sep=1pt,minimum width=1.73cm, minimum height=1.1cm]
		\tikzstyle{compilation_step}=[draw,rectangle,minimum width=0.4cm, minimum height=0.9cm,align=center,font=\scriptsize]
		\tikzstyle{compilation_stage}=[thick, dashed, rectangle, draw,inner sep=3pt,font=\scriptsize]
    	\tikzstyle{pass}=[draw, rectangle, dashed, font=\scriptsize,minimum width=0.2cm,minimum height=0.78cm]

		\node[compilation_step,dashed] (thinlink) at (1.1,0) {
		 Thin\\
		 Link
		};

	\begin{scope}[shift={(2.4,0)}]
		\newcommand{\optcg}{
		\node[compilation_step] (opt) at (.9,0) {Opt};
		\node[compilation_step,font=\tiny] (pre1) at (1.6, 0) {Pre\\
CG
};
		\node[compilation_step,font=\tiny] (cg1) at (2.38, 0) {CodeGen};
		\node[pass,blue](out1m) at (1.6,-0.01){};
		\node[pass](out1) at (2.73,-0.01){};
		}

		\optcg
		\node[profile2, fit=(opt)(cg1)](prf1) at (1.68, 0.2){};
		\node[profile2, fit=(opt)(cg1)](prf1) at (1.71, 0.1){};
		\node[profile2, fit=(opt)(cg1)](prf1) at (1.73, 0.0){};
		\optcg
	\end{scope}

		\node[compilation_step,rounded corners,font=\tiny] (gpt) at (2.15,.45) {
		 Global\\
		 Prefix\\
		 Tree
		};
		\node[compilation_step, blue, rounded corners, font=\tiny] (gmi) at (2.15,.-.45) {
		 Global \\
		 Merge\\
		 Info
		};
          \node[compilation_step,rounded corners,teal,thick,dashed,fit=(gpt)(gmi)] (read) {};

		\node[compilation_step] (link) at (6,0) {Linking};

		\node[rounded corners, draw, rectangle, minimum height=0.3cm, minimum width=2.5cm, text width=3cm,text centered,font=\scriptsize] (go2) at (4.22, -1.42) {Global Function Outliner};
		\node[rounded corners, draw, rectangle, minimum height=0.3cm, minimum width=2.5cm, text width=3cm,text centered,font=\scriptsize, blue] (gm2) at (4.22, -1.9) {Global Function Merger};

\begin{pgflowlevelscope}{\pgftransformscale{.65}}
        \node[cloud, draw,fill=green!30, cloud puffs=20, cloud puff arc=120, aspect=1.4, inner ysep=0.05cm ] (cgdb) at (3.9,-1.5) {Artifacts};
\end{pgflowlevelscope}

	    \draw[->,very thick, teal](.54, -.88)
		    to [out=0, in=238, looseness=0] node[font=\tiny, above] {read}(read);

		\draw[->,very thick,blue] (out1m.south)
		    to [out=-90, in=180, looseness=0] (4.0, -1.1)
		    to [out=0, in=-90, looseness=0] node[midway, above, font=\tiny] {merge} (2.4, -1.1)
		    to [out=270, in=90, looseness=0] (2.4, -1.9)
		    to (gm2.west);

		\draw[->,very thick] (out1.south)
		    to [out=90, in=0, looseness=0] (5.13, -1.1)
		    to [out=180, in=-90, looseness=0] node[midway, above, font=\tiny] {outline} (6, -1.1)
            to [out=-90, in=90, looseness=0] (6, -1.42)
            to (go2.east);

\end{scope}
\end{tikzpicture}

\caption{}\label{fig:write_read_cg}
\end{subfigure}

\begin{subfigure}[t]{\columnwidth}
\centering
\begin{tikzpicture}
  \node[single arrow, draw, fill=blue!30, shape border rotate=270, minimum height=3cm, text centered] (arrow) at (0,0) {CI};
  \node[single arrow, draw, fill=blue!30, shape border rotate=270, minimum height=3cm, text centered] (arrow) at (11,0) {CI};

  \tikzstyle{cg}=[draw,rounded corners,rectangle,minimum width=0.4cm, align=center,font=\scriptsize]
  \node[cg] (write1) at (2.2,.5){build-write-codegen-artifacts};
  \node[cg] (write2) at (2.2,-0.5){build-write-codegen-artifacts};

  \node[cg] (read1) at (8.8,.9){build-read-codegen-artifacts};
  \node[cg] (read2) at (8.8,.3){build-read-codegen-artifacts};
  \node[cg] (read3) at (8.8,-.3){build-read-codegen-artifacts};
  \node[cg] (read4) at (8.8,-.9){build-read-codegen-artifacts};

  \node[cloud,  fill=green!30,draw, cloud puffs=20, cloud puff arc=120, aspect=1.8, inner ysep=0.05cm ] (cloud) at (5.5,0) {Artifacts};

  \draw[->,very thick, teal](write1.east) to node[midway, above, font=\tiny]{write}(cloud.west);
  \draw[->,very thick, teal](write2.east) to node[midway, below, font=\tiny]{write}(cloud.west);

  \draw[->,very thick, teal](cloud.east) to node[midway, above, font=\tiny]{read}(read1.west);
  \draw[->,very thick, teal](cloud.east) to node[near start, right, font=\tiny]{read}(read2.west);
  \draw[->,very thick, teal](cloud.east) to node[near start, right, font=\tiny]{read}(read3.west);
  \draw[->,very thick, teal](cloud.east) to node[midway, below, font=\tiny]{read}(read4.west);

  \draw[-,dotted, thick](.3,.5) to (write1.west);
  \draw[-,dotted, thick](.3,-.5) to (write2.west);

  \draw[-,dotted, thick] (read1.east) to (10.7,.9);
  \draw[-,dotted, thick] (read2.east) to (10.7,.3);
  \draw[-,dotted, thick] (read3.east) to (10.7,-.3);
  \draw[-,dotted, thick] (read4.east) to (10.7,-.9);

\end{tikzpicture}
	\caption{}	\label{fig:CI}
\end{subfigure}
\caption{Overview of Single Pass Code Generation for Global Function Merger with Global Function Outliner.
By leveraging previously generated codegen artifacts (a), independent compilation becomes feasible, in continuous build environments (b) where source code continually evolves.}
\end{figure}

Figure \ref{fig:write_read_cg} illustrates how we divide Figure \ref{fig:globalmerger_overview} into two distinct builds.
One build is responsible for generating the \CodeGen artifacts, while the other build consumes these artifacts from a prior build, eliminating the need for the initial \CodeGen pass.
As discussed in Section \ref{sec:soundness}, this separation is feasible because our transformation is safe even in the presence of potential changes in the IR and potential staleness in the \CodeGen artifacts.

Many large projects incorporate continuous integration (CI), which automatically triggers a full build at specific intervals.
In Figure \ref{fig:CI}, we first created a new CI build, \textit{build-write-codegen-artifacts}, which corresponds to the first \CodeGen step that publishes the \CodeGen artifacts.
This CI build can be triggered infrequently to refresh the \CodeGen artifacts.
Following this, we established a second CI build, \textit{build-read-codegen-artifacts}, which is more frequently triggered due to its role in conventional development and release processes.
\textit{build-read-codegen-artifacts} reads the \CodeGen artifacts generated from \textit{build-write-codegen-artifacts} that ran earlier, and feeds them into the \CodeGen step, which is only executed once.
\textit{build-read-codegen-artifacts} runs faster than the build described in Section \ref{sec:algorithm} because it runs the \CodeGen step only once instead of twice.
However, we may miss some opportunities for size savings because changes in the source code between when \textit{build-write-codegen-artifacts} ran and when \textit{build-read-codegen-artifacts} ran will affect the accuracy of the \CodeGen artifacts.  
Section \ref{sec:tradeoff} will provide insights into the impact on app size as the source code evolves, and the reduction in build time achieved when using this approach.


It is worth noting, as illustrated in Figure \ref{fig:write_read_cg}, that the ThinLink step is optional, allowing this single \CodeGen mode to remain independent of ThinLTO.
As each module compilation can be executed independently, this build scheme is  applicable to both NoLTO and even distributed ThinLTO on multiple machines.
Additionally, we emphasize that the presence of \CodeGen artifacts from either \GlobalPrefixTree or \GlobalMergeInfo is also optional. This means that either \gfo or \gfm can safely operate independently.
Looking ahead, we aim to expand this shared \CodeGen artifact framework to accommodate other optimistic size optimization techniques in the future.

\subsection{Interaction between function merger and function outliner}\label{sec:interaction}

Recall that \gfm is implemented at the IR level to handle similar functions, enabling easy parameterization without concerns about lowering them.
In contrast, \gfo is implemented at the MIR level, serving as a post-register allocation pass for efficiently outlining identical code sequences while minimizing call overhead.
These two passes not only operate at different levels but also work independently, each maintaining its own global summaries.

There are two potential interactions between these two passes.
First, the code transformation performed by \gfm can influence the efficiency of \gfo.
As we will illustrate in Section \ref{sec:tradeoff}, when considering the impact of changes in the source code over time, it becomes evident that outlining opportunities for \gfo remain relatively stable in the face of changes in the IR.
This resilience is due to the frequent occurrence of short, repetitive code sequences throughout the application binary.

Secondly, the code transformation carried out by \gfo can affect the merging candidates of \gfm.
For instance, when \gfm creates two merging instances, $f1.Tgm$ and $f2.Tgm$ from $f1$ and $f2$, respectively, the local outlining heuristic in \gfo may independently outline different piece of code sequences among neighboring functions within each module.
This process may result in alterations to those merging instances, $f1.Tgm$ and $f2.Tgm$, in a way that prevents the linker from folding them.

To address this issue, we made a slight modification to \gfo's heuristic.
Specifically, all merged functions (indicated by the suffix $.Tgm$) no longer undergo local outlining heuristic, but are matched only with the \GlobalPrefixTree to find outlining candidates.
This adjustment allows us to maintain deterministic outlining for the merged functions across modules, significantly reducing mismatches.

\subsection{Improving hashes for Objective-C Globals}\label{sec:objcglobal}

While the precision of the hash does not impact the correctness of \gfm, it can significantly influence the efficiency of the merging process.
If the hash is overly conservative, it may group distinct functions, leading to the unnecessary creation of merging candidates and resulting in code bloat. On the other hand, if we treat all private globals as distinct, we miss the opportunity to group similar functions that ultimately point to the same global.

When computing stable hashes for globals, we do not include hashes of pointers.
Instead, we expand data structures before applying the hash. This method ensures that the hashes remain unaffected by the non-deterministic placements of data structures in memory, preserving their meaning across modules.
For public globals, we hash their names, assuming that globals with the same name are likely to be identical.

In the case of Objective-C, the compiler emits various types of global metadata to enable runtime introspection.
These globals are often private and have unique name sequences.
Instead of hashing these private names, we hash the contents of their internal data structures which are meaningful across modules.
For instance, we hash the string contents of string literals located in sections like \textit{\_\_cstring} and \textit{\_\_objc\_methname}.
Another example is a selector reference global located in section \textit{\_\_objc\_selrefs} whose contents is another global reference to \textit{\_\_objc\_methname}.
In this case, we hash the contents of the referenced string literals.



\subsection{Debugging}\label{sec:debug}
When two functions are merged, a single merged function represents what were originally two distinct functions, with only one function's debug information surviving.
This means, at a given binary location, debug information maps to a single source location.
While there was a proposal \cite{dwarfmerge} to address this limitation in debug information,
it does not appear to have been realized.
Unlike traditional function merging instances that might be optimized away by the compiler,
\gfm effectively relies on the linker to fold the merging instances that remain identical.
Therefore, all changes related to debugging are similar to typical instances of ICF by the linker.

Fortunately, the linker produces a linker map to indicate which functions have actually been folded.
With this information, we can infer what a given stack trace may actually represent.
In a production environment where crash reports are published with stack traces,
developers often examine call stacks to understand what might have occurred.
However, unexpected functions in the stack traces (due to function merging) can confuse them.
By using this linker map, we can decorate the merged functions within a specific stack trace.
This enables developers to consider the presence of aliased functions at a particular frame and cross-reference potential functions from the linker map.
An ideal support tool could leverage a call-graph to prioritize potential functions by matching the parent (caller) frame in the call-graph.

An alternative approach involves creating a stack frame whenever a thunk is generated, rather than forming a tail-call. However, this method results in a larger size overhead for the thunk, which reduces the benefits of function merging. Despite this drawback, it may be justifiable for critical components where preserving debug information is essential.

\section{Evaluation}\label{sec:evaluation}
In Section \ref{sec:setup}, we introduce benchmarks and environments.
In Section \ref{sec:codesize}, we compare code size improvements with function outliners and mergers.
Section \ref{sec:quality} presents various merging quality statistics.
Section \ref{sec:tradeoff} discusses build-time and app size.
Finally, Section \ref{sec:perf} briefly covers performance insights.

\begin{table}[!tb]
  \caption{Statistics of Applications used for Evaluation.}
  \label{tab:benchmark}
  \centering
  \begin{tabular}{@{} lrrcc @{}} 
      \toprule
  App & File Size & Functions & Language & OS  \\
          & (MB) & (K) &   &         \\
  \midrule
  \fb &  233 & 2712 & Obj-C/Swift & iOS \\
  \bizapp &  143 & 1877 & Obj-C/Swift & iOS \\
  \fbatwork &  124 & 1506 & Obj-C/Swift & iOS \\
  \msg & 76 & 1020 & Obj-C/C++ & iOS \\
  \talk & 81 & 1001 & Obj-C/C++ & iOS \\
     \midrule
  \textsf{gcc} & 10 & 33 & C & macOS\\
  \textsf{omnetpp} & 2 & 13 & C++ & macOS\\
  \textsf{xalancbmk} & 7 & 27 & C++ & macOS\\
  \textsf{parest} & 3 & 12 & C++ & macOS\\
  \textsf{blender} & 17 & 67 & C/C++ & macOS\\
  \bottomrule
  \end{tabular}
\end{table}

\subsection{Experimental Setup}\label{sec:setup}

As summarized in Table \ref{tab:benchmark},
We conducted an evaluation of our approach using five major real-world iOS apps.
\fb is one of the largest non-gaming mobile apps, featuring a multitude of dynamically loaded libraries (dylibs) that collectively contribute to a total binary size exceeding \SI{230}{MB}.
In contrast, \bizapp and \fbatwork are other large social apps, primarily catering to business users, with sizes smaller than that of \fb.
These apps are written using a blend of Objective-C and Swift, showcasing the versatility of our approach in optimizing code size in diverse language environments.
Additionally, \msg and \talk are medium-sized mobile apps, combining Objective-C and C++, and having a total size of approximately \SI{80}{MB}.
In addition to these mobile apps, we selected several large C/C++ benchmarks from Spec CPU\textregistered{} 2017 \cite{speccpu2017} for comparative purposes, although they may not be representative of typical mobile app characteristics.
These benchmarks remain relatively small when compared to our real-world mobile apps, which range in size from \SI{2}{MB} to \SI{17}{MB}.
Throughout our evaluation, all benchmarks were built with the minimum size \textit{-Oz} optimization flag under ThinLTO, aiming to produce minimal code with reasonable build performance.

\subsection{Code Size}\label{sec:codesize}
\begin{figure}[!tb]
  \centering
  \begin{subfigure}[t]{0.49\columnwidth}
  \includegraphics[width=\linewidth]{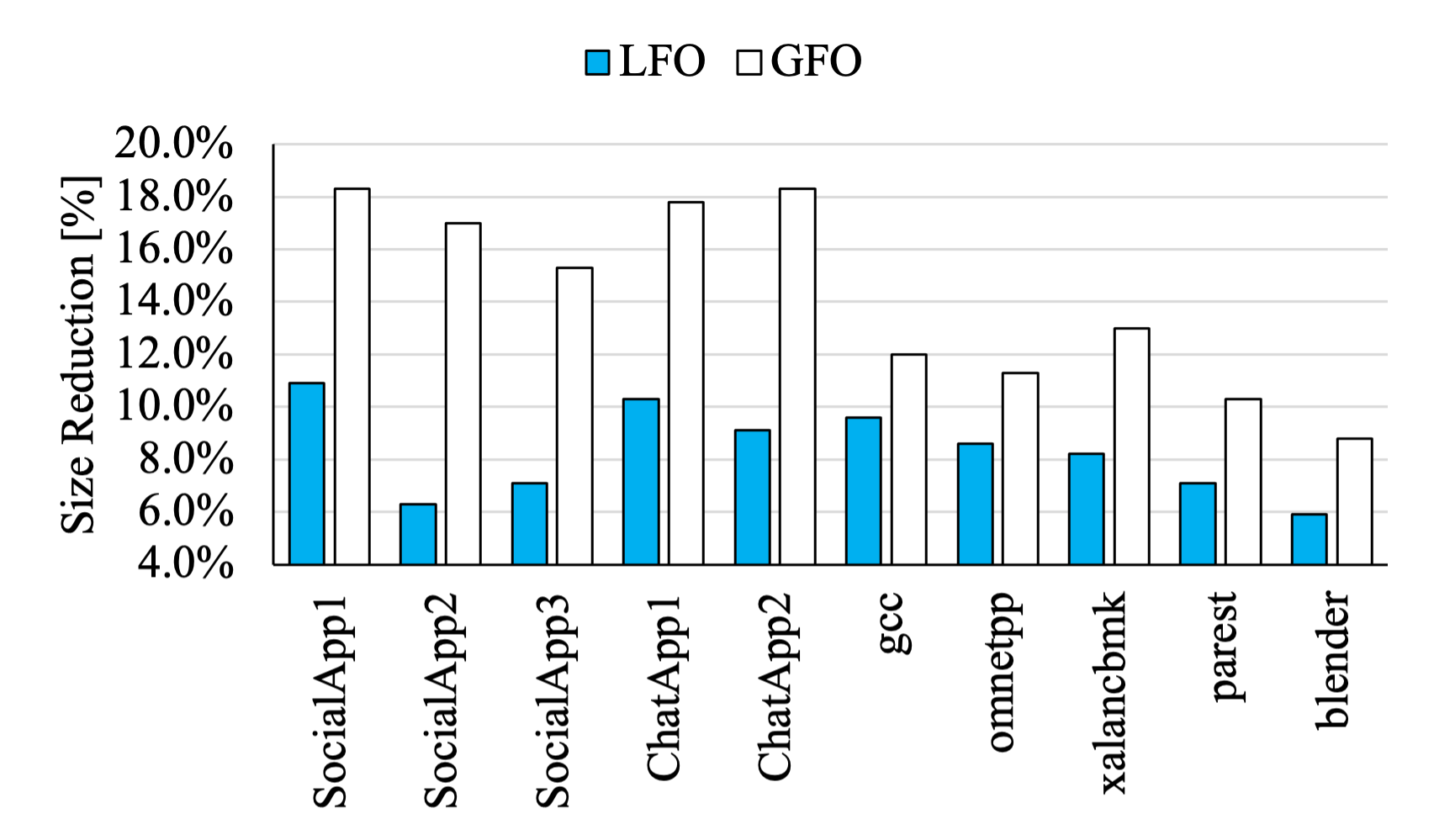}
  \caption{}\label{fig:outline-bar}
  \end{subfigure}
  \begin{subfigure}[t]{0.49\columnwidth}
  \includegraphics[width=\linewidth]{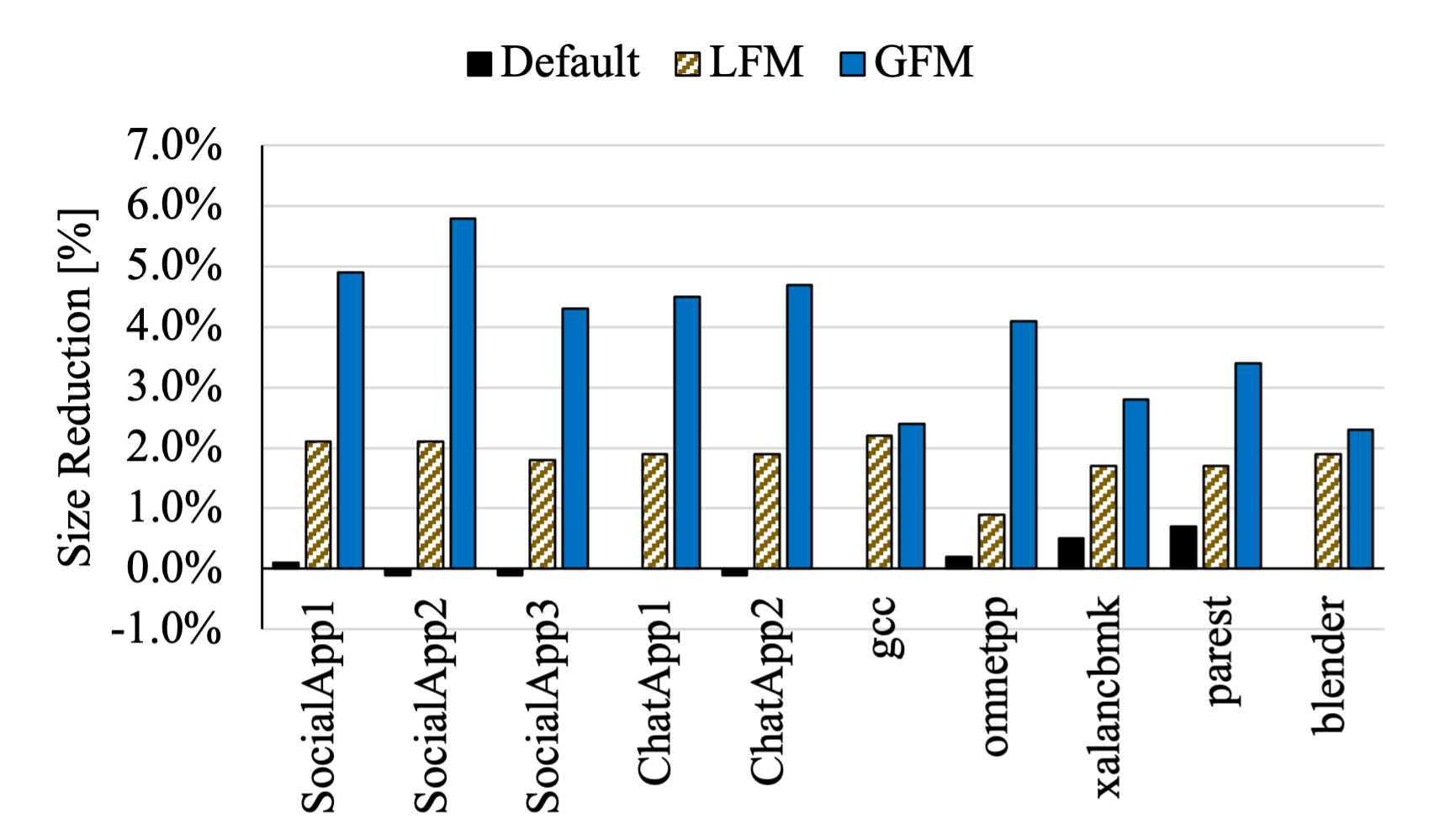}
  \caption{}\label{fig:merge-bar}
  \end{subfigure}
  \begin{subfigure}[t]{0.65\columnwidth}
  \includegraphics[width=\linewidth]{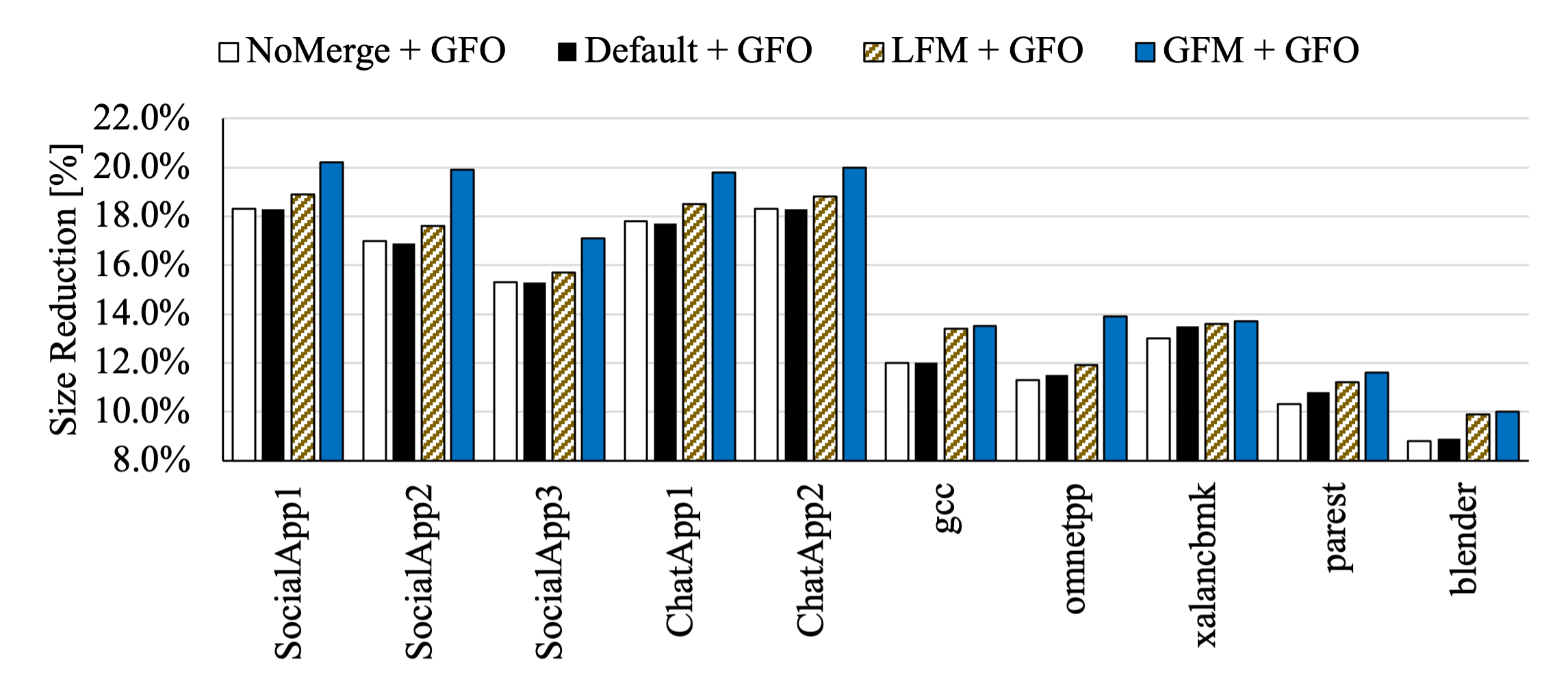}
  \caption{}\label{fig:merge-outline-bar2}
  \end{subfigure}
\caption{(a)~Size Improvement using Function Outliners without Function Merges, (b)~Size Improvement using Function Mergers without Function Outliners, and (c)~Size Improvement using Function Mergers with \gfo. }\label{fig:outline-merge-bar2}
\end{figure}

Figure \ref{fig:outline-bar} illustrates the code size improvements achieved by the default local function outliner (\lfo) \cite{machineoutliner} and the global function outliner (\gfo), as described in Section\ref{sec:globaloutliner}.
The size reduction achieved by \lfo ranges from 6\% to 10\% across the benchmarks.
However, the code size reduction with \gfo is significantly more pronounced in larger mobile apps compared to the relatively smaller Spec benchmarks.

In Figure \ref{fig:merge-bar}, we present the code size improvements using three different function merging techniques: the default technique available in LLVM (\default), the local function merger we integrated into LLVM from Swift's function merger (\lfm), and the global function merger (\gfm).
All outlining techniques are disabled in this comparison.
Expectedly, \default approach does not lead to significant size reductions, as the linker's ICF already performs a similar task.
\lfm results in a size reduction ranging from 1\% to 2\% across the benchmarks.
Notably, \gfm proves to be highly effective for larger binaries. In particular, \bizapp demonstrates the most substantial reduction, reaching up to 5.8\%.
Comparing \bizapp with \fb that consists of numerous smaller binaries, \bizapp  has a single large binary, offering more opportunities for merging within a single, large scope.

In Figure \ref{fig:merge-outline-bar2}, we enable these function merging techniques alongside state-of-the-art \gfo.
As the function outliners effectively extract identical code sequences, the benefits from function merging shrink as the body of the merged functions themselves can be outlined further.
Nevertheless, \gfm stands out by significantly enhancing the size reduction beyond the initial gains from \gfo.
In fact, most mobile apps achieve reductions exceeding 20\% when both techniques are applied together while more modest size reductions are observed in the small Spec benchmarks.

\subsection{Merging Quality}\label{sec:quality}

This section shows various statistics to assess merging quality with mobile apps.

\paragraph*{Mismatch Rate.}
Table \ref{table:mismatch} presents the percentage of merged function count and mismatched function count relative to the total function count.
Merged functions refers to those that were optimistically created using \GlobalMergeInfo.
On the other hand, mismatched functions are those among the merged functions that were not folded via the linker's ICF.
In such cases, the original functions unnecessarily become thunks that call the unique merging instance, resulting in unnecessary code size bloat.
This occurs because our merge transformation is based on hash-based summaries, and some mismatched cases are inevitable.

Overall, a large portion of functions, approximately 15\%, is attempted to be merged, while the number of mismatched functions within the total function count remains relatively low, at only 1\%.
The ratio of mismatched to merged functions provides another interesting insight.
\bizapp shows the lowest ratio at 2.46\%.
This means that, given the merging attempt, this application demonstrates the most efficient merging.
This result aligns with why \bizapp exhibits the largest relative size improvement from \gfm, as shown in Figure \ref{fig:merge-outline-bar2}.

\begin{table}[!tb]
	\small
	\centering
	\caption{Merged or Mismatched Functions Relative to the Total Functions.}
	\label{table:mismatch}
	\begin{tabular}{lrrrrr}
		\toprule
		\centering
		& \fb & \bizapp & \fbatwork & \msg & \talk \\
		\midrule
		\textsf{Merged} &  14.38\% & 15.92\% & 13.61\% & 13.09\% & 14.04\% \\
		\textsf{Mismatched} & 0.53\% & 0.39\% & 0.66\% & 0.46\% & 0.76\% \\
		\midrule
		\textsf{Mismatched/Merged} & 3.67\% & 2.46\% & 4.84\% & 3.55\% & 5.44\%\\
		\bottomrule
	\end{tabular}
\end{table}

\begin{figure}[!tb]
  \centering
  \begin{subfigure}[t]{0.49\columnwidth}
  \includegraphics[width=\linewidth]{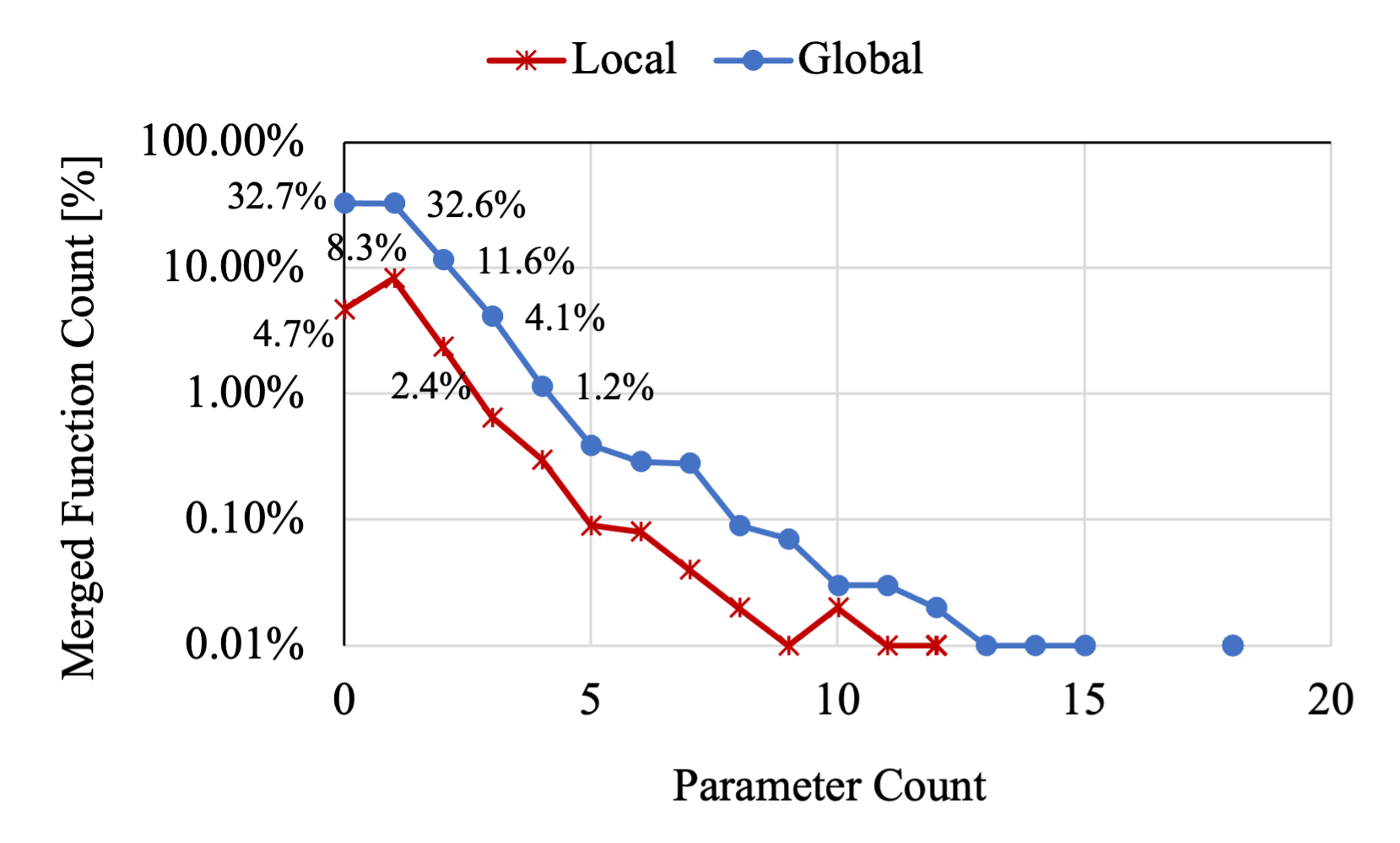}
  \caption{}\label{fig:params}
  \end{subfigure}
  \begin{subfigure}[t]{0.49\columnwidth}
  \includegraphics[width=\linewidth]{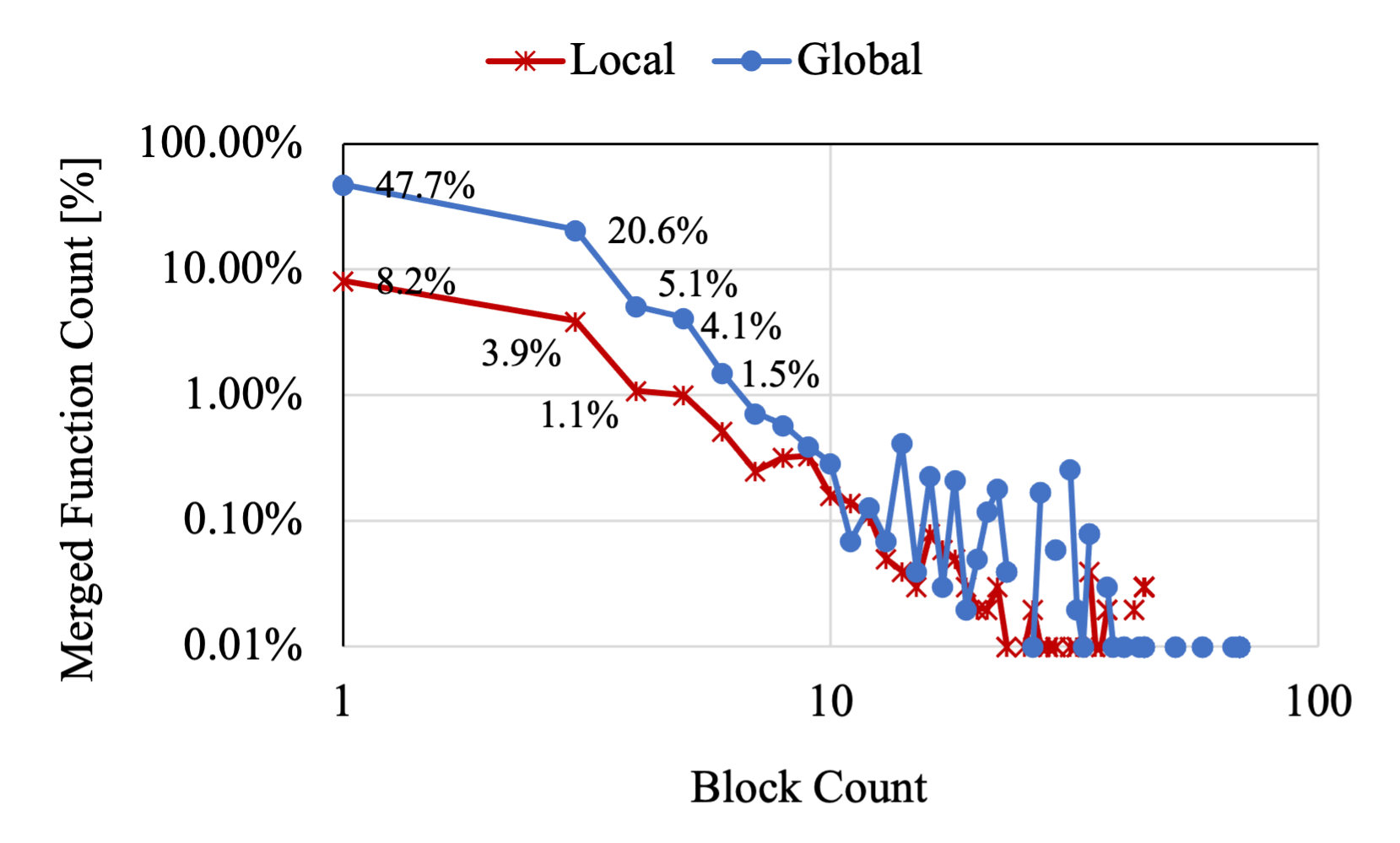}
  \caption{}\label{fig:blocks}
  \end{subfigure}
\caption{Percentage of merged function count with varying parameter count and block count.}\label{fig:params-blocks}
\end{figure}

\paragraph*{Parameters and Blocks.}
Figure \ref{fig:params-blocks} shows the percentage of merged function count with varying parameter count and block count with \fb.
In this figure, \textit{Local} means the merged functions within a module or compilation unit while \textit{Global} means those across modules.
Overall, 17\% of merged functions are \textit{Local} while 83\% of them are \textit{Global}.

Over 70\% of merged functions have either 0 or 1 parameter.
When the parameter count exceeds 4, the number of remaining merged functions drops to less than 1\%.
The average parameter count is 0.99, with a standard deviation of 1.21.
Similarly, more than 70\% of merged functions consist of up to 3 basic blocks.
When the block count exceeds 6, the number of remaining merged functions drops to less than 1\%.
However, the block count exhibits higher variability.
The average block count is 2.90, with a standard deviation of 4.66.

\begin{figure}[!tb]
  \centering
  \begin{subfigure}[t]{0.49\columnwidth}
  \includegraphics[width=\linewidth]{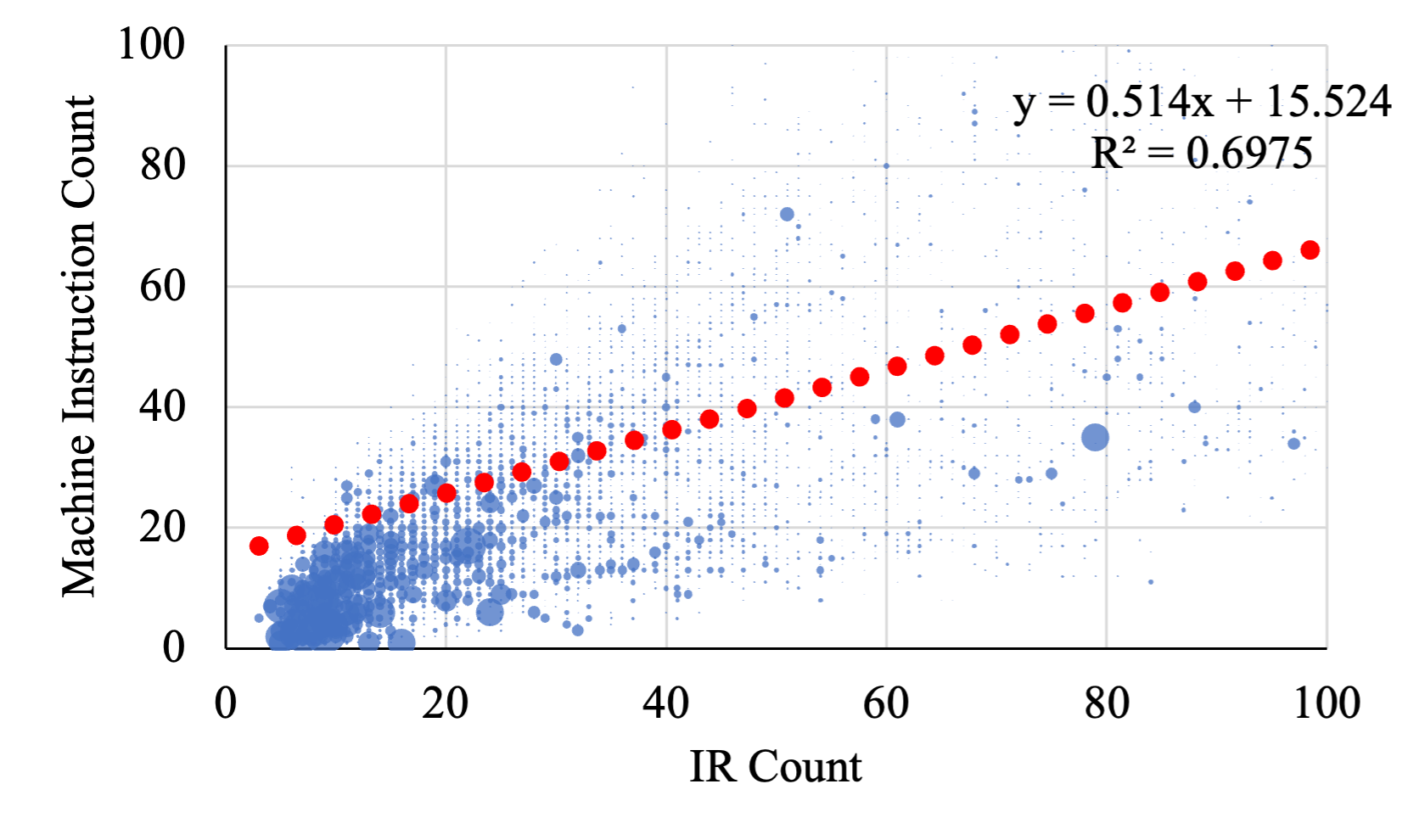}
  \caption{Without function outliners}\label{fig:size-nooutline}
  \end{subfigure}
  \begin{subfigure}[t]{0.49\columnwidth}
  \includegraphics[width=\linewidth]{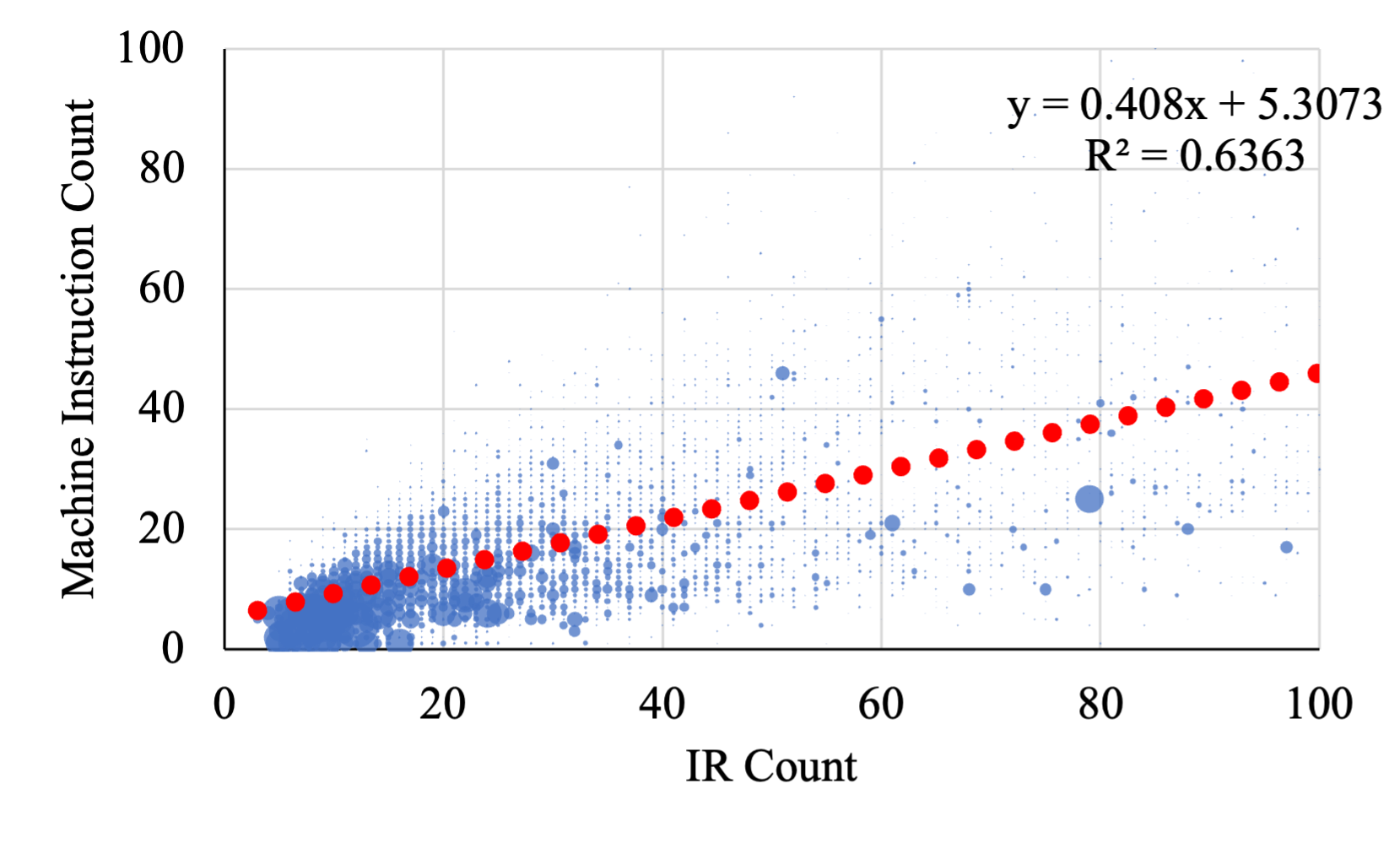}
  \caption{With function outliners}\label{fig:size-outline}
  \end{subfigure}
\caption{IR Count vs. Machine Instruction Count for Merged functions.  Frequent data points create large, closely clustered blue bubbles.}\label{fig:size-bubbles}
\end{figure}

\paragraph*{Correlation between IR and Machine Instruction.}
Figure \ref{fig:size-bubbles} illustrates the correlation between the IR count and the machine instruction count for merged functions with \fb targeted for arm64  architecture \cite{arm64} where each machine instruction has 4 byte size.
With function outliners disabled, the observed trend follows a moderate linear correlation, as indicated by the equation $y=0.514x+15.524$, with an associated R-squared value of $R^2 =0.6975$.
This correlation depends on how a single IR can either expand into multiple machine instructions or how multiple IRs can be optimized into fewer machine instructions.
It's influenced by the semantics and the quality of subsequent optimizations applied to the IR.

With function outliners enabled, the machine instruction count per IR tends to decrease due to further size reduction from outlining.
However, the linear trend becomes less pronounced, described by the equation $y=0.408x+5.3073$, with an associated R-squared value of $R^2 = 0.6363$.
Since outlining can occur in diverse ways, estimating the machine instruction count for a given IR becomes more challenging when function outliners are considered.
In other words, modeling size costs at the IR level with function merging when function outliners are enabled presents a significant challenge.
Future work may focus on improving this area, potentially leveraging artificial intelligence (AI) \cite{mlgo,cummins2023large} to address these complexities.

\paragraph*{Mergeability by Language Feature.}

As presented in Table \ref{table:lang}, we have categorized all functions into their respective language feature groups and calculated the $mergeability$ for \fb.
$mergeability$ for each language feature is determined by dividing the size of merged functions by the size of all functions within that specific language feature.
Notably, C++ templates exhibit a $mergeability$ of over 21\%, followed by Objective-C blocks and Swift functions.
In this application, the UI framework \cite{componentkit} prominently relies on C++ templates, while Swift UI plays a relatively smaller role.
It's worth noting that as Swift code continues to grow, the landscape may evolve in the future.

\begin{table}[!tb]
	\small
	\centering
	\caption{Mergeability per Language Category for \fb.}
	\label{table:lang}
	\begin{tabular}{lr}
		\toprule
		\centering
            & mergeability \\
            \midrule
		\textsf{Templates in C++} & 21.7\% \\
		\textsf{Objective-C block related} & 16.1\% \\
            \textsf{Swift} & 12.5\% \\
		\textsf{Objective-C (excluding block related)} & 10.0\% \\
		\textsf{C++ (excluding templates)} & 5.4\% \\
		\textsf{Other} & 4.4\% \\
		\bottomrule
	\end{tabular}
\end{table}

\subsection{Build Time and App Size}\label{sec:tradeoff}

In Section \ref{sec:singlecg}, we discussed a single \CodeGen alternative that improves build time at the cost of app size.
In this section, we present experimental data using \fb.
We established a CI build named \textit{build-write-codegen-artifacts} which generates \CodeGen artifacts and stores them on our servers weekly.
Additionally, we set up another CI build named \textit{build-read-codegen-artifacts} that reads the most recently published \CodeGen artifacts.
We conducted tests with two configurations of these builds: one using \gfo only with \GlobalPrefixTree as \CodeGen artifacts, and the other employing \gfm in conjunction with \gfo, using \GlobalMergeInfo and \GlobalPrefixTree as \CodeGen artifacts.

\begin{figure}[!tb]
  \centering
  \begin{subfigure}[t]{0.49\columnwidth}
  \includegraphics[width=\linewidth]{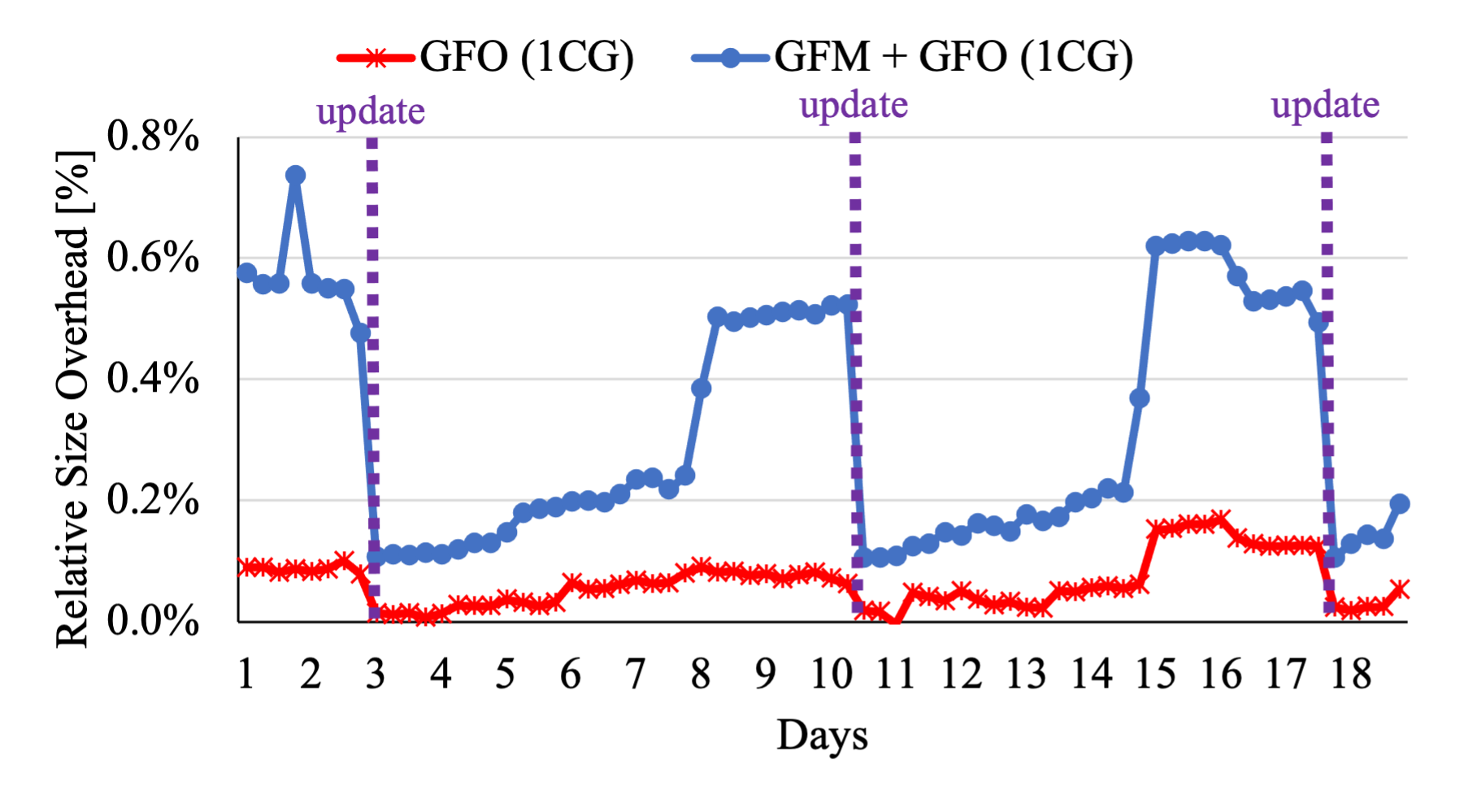}
  \caption{}\label{fig:overhead}
  \end{subfigure}
  \begin{subfigure}[t]{0.49\columnwidth}
  \includegraphics[width=\linewidth]{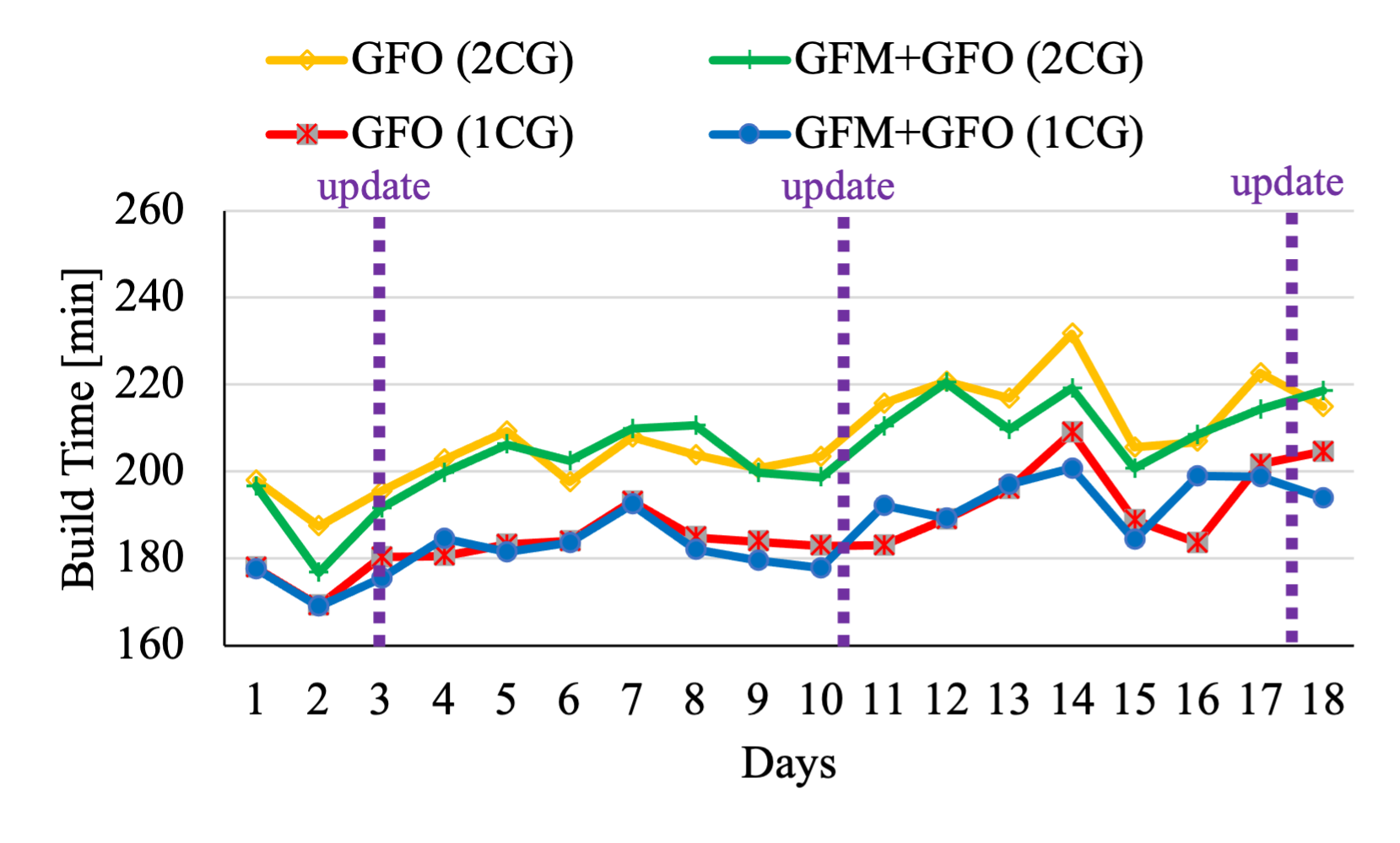}
  \caption{}\label{fig:buildtime}
  \end{subfigure}
\caption{(a)~App Size Overhead with Single \CodeGen Relative to Two-Round \CodeGen, and (b)~Build Time with Single \CodeGen and Two-Round \CodeGen. This data was collected over nearly three weeks of continuous builds.}\label{fig:buildtime-overhead}
\end{figure}

Figure \ref{fig:overhead} presents the (uncompressed) app size increase relative to the baseline of two-round \CodeGen for \textit{build-read-codegen-artifacts} that runs one-round \CodeGen.
We monitored these sizes over the past three weeks, during which there were typically hundreds or even thousands of daily commits to the repository although some of these commits are unrelated to the actual source changes for the app.
The app size increase from a stale summery with \gfo is minimal, less than 0.2\%, even when there are many source code changes during the week.
This is because the outlining opportunities primarily involve short, repetitive code segments that remain relatively stable across the app binary.
However, when \gfm is enabled in conjunction with \gfo, the app size increase becomes more significant.
Just before the purple vertical lines, the size increase spikes to as high as 0.7\% due to outdated artifacts.
Immediately after we refresh the artifacts, the increase drops to nearly 0.1\%.

Figure \ref{fig:buildtime} illustrates the end-to-end build time for these scenarios. When comparing single-round \CodeGen (referred to as 1CG) with two-round \CodeGen (referred to as 2CG), the average build time for \textit{build-read-codegen-artifacts} is, on average, 9\% faster.
A comparison between \mbox{\gfm+\gfo} and \gfo allows us to evaluate the additional build time introduced by \gfm.
This difference falls within the margin of error, suggesting that the extra build time incurred by \gfm is minimal, regardless of whether one-round or two-round \CodeGen is employed.

\subsection{Performance}\label{sec:perf}
\begin{table}[!tb]
	\small
	\centering
	\caption{The Relative Performance of Selected CPU Speed Benchmarks When Using \gfo or \gfm.}
	\label{table:perf}
	\begin{tabular}{crrrrrr}
		\toprule
		\centering

		& \multicolumn{2}{c}{\textsf{gcc}}
		& \multicolumn{2}{c}{\textsf{omnetpp}}
        & \multicolumn{2}{c}{\textsf{xalancbmk}} \\

            & \multicolumn{1}{c}{Avg} & \multicolumn{1}{c}{StDev}
            & \multicolumn{1}{c}{Avg} & \multicolumn{1}{c}{StDev}
            & \multicolumn{1}{c}{Avg} & \multicolumn{1}{c}{StDev} \\
            \midrule
		\gfo & 98.9\%	& 1.0\% &	96.1\% & 0.4\% &	77.2\%	& 1.1\% \\
		\gfm+\gfo & 93.1\% &  0.7\% &	96.3\% & 0.2\% &	 77.8\%	& 0.9\% \\
		\bottomrule
	\end{tabular}
\end{table}

Function merging and function outlining are designed for size optimization at the cost of additional call overhead, which can impact performance.
However, these aggressive size optimization techniques generally do not impede the performance of mobile apps that are primarily I/O-bound.
Our findings align with previous research \cite{uber,LFZLS22}.
For CPU-bound scenarios such as scroll performance or message sending in mobile apps, we employ profile-guided optimization (PGO) \cite{pgo-mobile} to selectively apply these optimizations only to cold functions.

While we do not recommend using these techniques without PGO for performance-critical scenarios like typical server workloads,
Table \ref{table:perf} provides insights into the expected performance impact using a subset of CPU speed benchmarks conducted over 5 iterations.
The baseline is the default -Oz compilation at 100\%.
Values below 100\% indicate performance regression.
As expected, \gfo leads to a significant performance slowdown in CPU-bound cases, up to 23\% with \textsf{xalancbmk}.


When \gfm is applied in addition to \gfo, \textsf{gcc} experiences an additional 5.8\% slowdown, reducing its performance from 98.9\% to 93.1\%.
The presence of thunks and higher register pressure resulting from the merged functions impacts the efficiency of the hot code path.
Interestingly, \textsf{omnetpp} and \textsf{xalancbmk} exhibit a modest improvement, although their performance still falls within the standard deviation range. This suggests that function merging can consolidate similar functions into a single location, potentially eliminating numerous small, scattered outlined call-sites.
This consolidation may enhance the instruction cache (I-cache) locality, which aligns with previous work \cite{simfunc}.







\section{Related Work}\label{sec:relatedwork}
\paragraph*{Function Merging.}
While the existing function merger in LLVM, as described in \cite{mergefunc},
primarily merges identical functions at the IR level, the work presented in \cite{simfunc} extends this capability to handle similar functions sharing the same control-flow and an equivalent number of instructions.
Further advancements, as seen in \cite{SeqAlign2019, ssafuncmerge}, employ sequence alignment algorithms to merge arbitrary pairs of functions, followed by more enhanced versions \cite{hyfm, ssafuncmerge} to improve the efficiency of sequence alignment operations.
More recently, they have proposed HyBF \cite{HyBF2023}, which focuses on code merging techniques targeting conditional branches with similar code regions.

All of these prior approaches introduce additional control flows to construct new merged functions, potentially impacting the integrity of functions unless all metadata emitted by the front-end, such as Swift, is carefully considered.
More critically, none of these approaches propose a scalable build approach applicable to NoLTO or ThinLTO as they require explicit IR comparisons, which can be prohibitively expensive when building large applications.

Similar to the Swift function merge approach, as discussed in \cite{swiftmerge}, our method addresses nearly identical functions that differ only by constants.
These variations are particularly prevalent in high-level constructs such as C++ templates, Objective-C blocks, and Swift generated functions, making them particularly valuable for commercial mobile apps.
Functions that exhibit significant differences while sharing common code sequences are already efficiently managed by function outliners.
We firmly believe that our approach, thoughtfully integrating both \gfo and \gfm, offers a practical solution to reduce app size without incurring high build-time cost.



\paragraph*{Cross-Module Summary.}
In ThinLTO \cite{thinlto}, the combined bitcode summary, containing symbol information and references across modules, enables function inlining and function importing in each thread independently.
Expanding the bitcode summary to include function shapes is theoretically possible for function merging, but concurrent function inlining and function merging can lead to conflicts.
Function inlining may alter function shapes, rendering the summary unsuitable for function merging, and function merging can disrupt the function inlining process.
As discussed in Section \ref{sec:background}, it's more effective to apply function merging or function outlining later in the optimization pass when function shapes have stabilized.
Our \GlobalMergeInfo and \GlobalPrefixTree capture this information to support subsequent function merging and function outlining.

\paragraph*{Optimization Profile.}
The \CodeGen artifacts we generate can conceptually be thought of as an optimization profile, similar to the profiles used in traditional PGO frameworks \cite{bolt,SKR23,HLMP23}.
While PGO profiles capture the runtime behavior of a program, \CodeGen artifacts capture its behavior during the build process.
Both PGO profiles and \CodeGen artifacts are used at build-time to enhance the resulting binary.
PGO profiles identify performance-critical program segments, aiding optimizations like function inlining \cite{inlinerflow} and function layout \cite{ottoni} to improve code locality.
In contrast, \CodeGen artifacts identify similar program segments, which we use to optimize binary size through function outlining and function merging.

\paragraph*{Optimistic Transformation.}
Optimistic transformations include loop specialization \cite{OHJ13,loopopt} and indirect call promotion \cite{autofdo}.
To maintain the safety of these optimizations, it is necessary to incorporate additional runtime checks or fallback code.
In contrast, our optimistic approach restructures similar functions into a canonicalized form, ensuring safety guarantees by design.


\section{Conclusion}\label{sec:conclusion}

In this paper, we introduce a novel framework for efficient function merging and outlining that seamlessly operates across modules. Our approach has proven to significantly reduce the code size of real-world iOS apps, achieving an additional 3.5\% reduction through function merging, on top of the initial average reduction of 17.3\% accomplished through function outlining.
This results in an overall improvement of 20\%.

Unlike existing function mergers, which require processing entire modules through LTO and result in lengthy build times,
our methodology leverages global summaries to optimistically create merging instances in parallel.
Our approach ensures soundness even in the presence of changes in the IR, rendering it practical for large-scale app development in distributed build environments.

We have future plans to expand this work beyond handling various constants to dealing with different types.
We aim to investigate the finer-grained grouping of similar functions, which will demand more precise and possibly AI-driven cost modeling.
Additionally, we intend to apply this optimistic size optimization framework to other domains, including register allocation.
\newpage


\bibliographystyle{ACM-Reference-Format}
\bibliography{refs}

\end{document}